\documentclass[twocolumn,tighten]{aastex631}
\usepackage{apjfonts} 
\usepackage{amsmath,amstext}
\usepackage{graphicx}
\usepackage{rotating}
\usepackage{multirow}
\usepackage{verbatim}
\usepackage{threeparttable}

\begin{document}

\title{\LARGE Searching for rapid pulsations in solar flare X-ray data}
\author[0000-0003-0656-2437]{Andrew R. Inglis}
\affiliation{Solar Physics Laboratory, Heliophysics Science Division, NASA Goddard Space Flight Center, Greenbelt, MD, USA}
\affiliation{Physics Department, The Catholic University of America, Washington, DC, USA}
\author[0000-0002-6835-2390]{Laura A. Hayes}
\affiliation{European Space Agency (ESA), European Space Research and Technology Centre (ESTEC), Keplerlaan 1, 2201 AZ Noordwijk, The Netherlands}

\begin{abstract}
Most studies of quasi-periodic pulsations in solar flares have identified characteristic periods in the 5 -- 300s range. Due to observational limitations there have been few attempts to probe the < 5s period regime and understand the prevalence of such short-period quasi-periodic pulsations. However, the Fermi Gamma-ray Burst Monitor (GBM) has observed approximately 1500 solar flares to date in high cadence 16 Hz burst mode, providing us with an opportunity to study short-period quasi-periodic pulsations at X-ray energies. We systematically analyse every solar flare observed by Fermi/GBM in burst mode, estimating the prevalence of quasi-periodic pulsations in multiple X-ray energy bands. To better understand these results, we complement this with analysis of synthetic solar flare lightcurves, both with and without oscillatory signals present. Using these synthetic lightcurves, we can understand the likely false alarm and true positive rates in the real solar GBM data. We do not find strong evidence for widespread short-period quasi-periodic pulsations, indicating either a low base occurrence rate of such signatures or that their typical signal-to-noise ratios must be low -- less than 1 -- in Fermi/GBM data. Finally, we present a selection of the most interesting potential quasi-periodic pulsation events that were identified in the GBM solar X-ray data.

\end{abstract}


\section{Introduction}
\label{introduction}

Solar flares are rapidly evolving transient phenomena, capable of releasing up to 10$^{32}$ ergs of energy in a matter of minutes from the solar corona. Flares are efficient accelerators of energetic particles, and also heat the local solar atmosphere to millions of degrees. Thus, understanding flares is key for our knowledge of how energy is released in the solar atmosphere, and in highly magnetised plasma throughout the universe.

The time-variability of flare emission -- and periodic behaviour in particular -- is a subject of much interest. Quasi-periodic pulsations (QPPs) are a regularly observed phenomenon during both the impulsive and decay phases of flare emission \citep{2009SSRv..149..119N, 2016ApJ...833..284I, 2016SoPh..291.3143V, 2020ApJ...895...50H, 2021SSRv..217...66Z}. Although lacking a strict definition, the term QPPs typically refers to evidence of periodic behaviour in flare timeseries, though this behaviour may not be strictly oscillatory. Signals with evolving periods and anharmonic shapes are often included under the QPP umbrella. QPPs have been noted for over sixty years \citep[e.g.][]{1969ApJ...155L.117P, 1970SoPh...13..420C, 1978SoPh...57..191L} and can be observed throughout the electromagnetic spectrum, from radio waves and microwaves \citep[e.g.][]{2003ApJ...588.1163G, 2005A&A...439..727M, 2008A&A...487.1147I, 2013SoPh..284..559K, 2019NatCo..10.2276C}, to ultraviolet (UV) \citep[e.g.][]{2018ApJ...867...85B}, extreme-ultraviolet (EUV) \citep[e.g.][]{2018SoPh..293...61D}, X-ray \citep[e.g.][]{2001ApJ...562L.103A, 2012ApJ...749L..16D, 2015SoPh..290.3625S, 2020ApJ...895...50H, 2024arXiv240210546C}, and gamma-ray wavelengths \citep{2010ApJ...708L..47N}. In many cases, QPPs are a multi-wavelength feature of flare emission, seen similarly across many pass-bands \citep[e.g.][]{2021AAS...23830303C}. Similar quasi-periodic signatures have also been observed on other stars, primarily in the white light regime and occasionally in X-rays, for example \citet{2006A&A...456..323M, 2015ApJ...813L...5P, 2015MNRAS.450..956B, 2016MNRAS.tmp..660P, 2019A&A...629A.147B, 2021ApJ...923L..33K, 2022MNRAS.514.5178D}. QPPs are crucial to understand because they are signatures of fundamental physical processes that occur during solar flare energy release \citep{2018SSRv..214...45M, 2021SSRv..217...66Z}, including magnetic reconnection, particle acceleration, and magnetohydrodynamic (MHD) wave generation. A comprehensive understanding of QPPs therefore allows them to be used as potential diagnostic tools for understanding how energy is released in solar and stellar atmospheres.

Although QPPs have been detected with a wide range of periods, from sub-second \citep[e.g.][]{2020ApJ...903...63K} to tens of minutes \citep[e.g.][]{2005A&A...440L..59F}, due to observational constraints most detections of QPPs have historically been in the 5 - 300s range \citep[e.g.][]{2015SoPh..290.3625S, 2020ApJ...895...50H}. While it is well-established that solar X-ray emission can show structure at sub-second scales \citep[e.g.][]{1983ApJ...265L..99K,1984ApJ...287L.105K,1985SoPh..100..465D}, and case studies of rapid pulsations do exist \citep{2020ApJ...903...63K}, there have been few attempts to systematically search solar flare data for evidence of rapid quasi-periodic behaviour in the < 5s regime. Thus, the prevalence of rapid cadence QPPs in solar flares is currently unknown. This is crucial, since pulsations on such timescales imply the presence of periodic drivers of particle acceleration. In this work, we conduct one of the first such searches using high cadence data from the Fermi Gamma-ray Burst Monitor (GBM). Fermi is an astrophysics mission designed to observe distant transient events, but also observes solar flares. During transient events, the GBM often triggers a high cadence mode for certain binned data products, temporarily switching to 0.064s (16 Hz) temporal binning. We analyse all solar flares currently listed in the Fermi/GBM trigger catalog, a total of 1460 events. For each event, we search for evidence of QPPs in multiple energy ranges by fitting different models to the flare Fourier power spectra, following the methodology presented in \citet{2015ApJ...798..108I, 2016ApJ...833..284I, 2019ApJS..244...44B, 2020ApJ...895...50H, 2020JGRA..12527887M}. To better verify the results and understand the potential for false alarms, we complement this observational search with tests of the methodology on sets of simulated lightcurves.

In Section \ref{data_and_methods} we describe the data sources and the methodology used in this research. In Section \ref{simulations} we use sets of simulated lightcurves to evaluate the performance of our methodology and understand the potential false alarm rates. In Section \ref{gbm_data_analysis} we apply our methodology to the Fermi/GBM burst mode data from 1460 solar flares. We discuss the results of the statistical search for rapid pulsations in Section \ref{statistical_results}, and present a selection of the most interesting QPP events in Section \ref{case_studies}. The entire work is summarized in Section \ref{summary}.

\section{Data and methodology}
\label{data_and_methods}

\subsection{Fermi/GBM data}
For this work, we use data from the Fermi/GBM instrument \citep{2009ApJ...702..791M}. GBM consists of 12 NaI detectors and 2 BGO detectors, with their pointing axes arranged at different angles such that together they can observe the entire visible sky and localize transient astrophysical phenomena such as gamma-ray bursts (GRBs). The NaI detectors cover the energy range $\sim$ 4 keV -- 1 MeV, while the BGO detectors complement this with a higher energy range of $\sim$ 200 keV -- 40 MeV. 

GBM produces three main science data products; the CSPEC, CTIME, and TTE (time-tagged event) data products. TTE data provides a list of the individual detected photons each with timestamp information. CSPEC data provides binned count rates over time using 128 energy bins with a default temporal cadence of $\sim 4s$. The CTIME data product sacrifices energy resolution in favour of higher temporal resolution; it provides binned count rates in 8 energy bins with a default time resolution of 0.256s, and 0.064s in trigger mode.  

The GBM was designed to trigger different science operational modes when transient events -- such as GRBs or solar flares -- are detected. Since GBM is able to localize events, a solar flare can be identified when the source location is consistent with the position of the Sun. For the CTIME product, the time resolution is increased from 0.256s to 0.064s for a period of time after the trigger occurs, typically 600s, though it can be shorter. It is this high time resolution burst mode CTIME data that is the focus of this work, as it provides us with the opportunity to search for quasi-periodic signals in the 0.1s - 5s period regime.

We begin with the Fermi/GBM trigger catalog\footnote{\url{https://heasarc.gsfc.nasa.gov/W3Browse/fermi/fermigtrig.html}}, which at the time of analysis contained 1460 events marked as solar flares between 2008 to present. Fermi/GBM data has been previously analysed by \citet{2016ApJ...833..284I} at 1s cadence in search of quasi-periodic pulsations. In that work, the distribution of detected periods was centered at $\sim$ 10s, with periods of $<$ 5s largely inaccessible due to the chosen temporal cadence. The GBM burst mode data provides us with an opportunity to explore the $<$ 5s regime in search of pulsations. In this work, we limit our search for pulsations to this short period regime, while acknowledging that GBM is effective at detecting longer period pulsations in flares.

\subsection{Time series analysis with the AFINO tool}

To search for flare quasi-periodic pulsations with characteristic periods in the <5s regime, we use the AFINO \citep[Automated Flare Inference of Oscillations]{2015ApJ...798..108I, 2016ApJ...833..284I} analysis tool. We apply AFINO to GBM flare X-ray data in three binned energy ranges provided by the CTIME product, 4 - 15 keV, 15 - 25 keV, and 25 - 50 keV. AFINO has been previously described in detail by \citet{2015ApJ...798..108I, 2016ApJ...833..284I, 2018JGRA..123.6457M, 2020JGRA..12527887M, 2020ApJ...895...50H}, therefore we discuss only the key aspects of the method here. 

For a time series of interest, AFINO works by fitting a selection of models, one of which includes an oscillatory signal, to the Fourier power spectrum of the normalized, apodized original series. The original signal is not de-trended during this analysis, to avoid introducing artifacts into the results \citep[see][for a discussion of this issue]{2018SoPh..293...61D}. Normalization is done for convenience only, however the apodization -- that is, the multiplication of the original signal by a window function -- is important to mitigate the effects of having a non-infinite time series. In this work we use the Hanning window function. Following these steps, the Fourier power spectrum is fit by each model using a maximum likelihood method. Then, these models are compared using the Bayesian Information Criterion (BIC) to determine which model is most probable given the data. The BIC is defined as

\begin{equation}
BIC = - 2 \ln(L) + k \ln(n)
\label{bic_eqn}
\end{equation}

where $L$ is the maximum likelihood, $k$ is the number of free parameters and $n$ is the number of data points in the power spectrum. 

In this paper, we test three models of the Fourier power spectrum, one of which is representative of an oscillation in the original time series. These models may be written

\begin{equation}
S_0(f) = A_0 f^{-\alpha_0} + C_0
\label{m0_eqn}
\end{equation}

\begin{equation}
S_1(f) = A_1 f^{-\alpha_1} + B \ \exp \left( \frac{-(\ln f - \ln f_p)^2}{2\sigma^2} \right) + C_1
\label{m1_eqn}
\end{equation}

\begin{equation}
  S_2(f)=\begin{cases}
    A_2 f^{-\alpha_{b}} + C_2, & \text{if $f<f_{break}$}.\\
    A_2 f^{-\alpha_{b} - \alpha_{a}} f^{-\alpha_{a}} + C_2, & \text{if $f>f_{break}$}.
  \end{cases}
  \label{m2_eqn}
\end{equation}

for frequencies $f$, where $A_0$, $A_1$, $A_2$ and $B$ are amplitudes, $\alpha_0$, $\alpha_1$, $\alpha_a$ and $\alpha_b$ are power law indices, $C_0$, $C_1$, and $C_2$ are constants, and $f_p$ and $\sigma$ represent the location in frequency of a Gaussian peak and its width respectively. 

In descriptive terms, model $S_0$ is the simplest, a single power law plus a constant term. This model is motivated by the fact that power law power spectra are ubiquitous in astrophysical phenomena \citep{2005A&A...431..391V, 2010MNRAS.402..307V, 2010AJ....140..224C, 2011A&A...533A..61G, 2013ApJ...768...87H}. Model $S_1$ represents an oscillation in the data, including a localized Gaussian frequency enhancement at $f_p$ in addition to a power law and a constant term. In this work, we limit the maximum allowed period to 10s in order to focus on short-period events. The minimum period is determined by the Nyquist frequency. The width $\sigma$ of the Gaussian enhancement is also constrained to lie in the range 0.03 $< \log \sigma <$ 0.15, preventing the frequency enhancement from being unrealistically narrow or broad. Model $S_2$ is a broken power law, an alternative to model $S_0$ that can capture more complex features in the Fourier spectrum but does not contain an oscillation. 

For each fitted model, we calculate BIC and compare the QPP-like model with all other models, i.e. we calculate $\Delta$BIC = $BIC_j$ - $BIC_{QPP}$ for all the non-QPP models $j$. Since the fitting procedure minimizes $L$ and therefore $BIC$, we are interested in events where $\Delta$BIC is positive for all $j$. A typical threshold used to identify strong evidence in favour of one model over another is $\Delta$BIC > 10, as discussed in \citet{doi:10.1080/01621459.1995.10476572,Burnham01112004, 2015ApJ...798..108I,2016ApJ...833..284I,2019ApJS..244...44B,2020ApJ...895...50H}. When analysing a timeseries, when the oscillation model $S_1$ is strongly preferred to models $S_0$ and $S_2$ according to the $BIC$ comparison, we consider there to be strong evidence of a quasi-periodic signal in that timeseries with characteristic frequency $f_p$. The premise of this approach therefore is that a quasi-periodic signal can be adequately represented by a local frequency enhancement superimposed on a power-law background in the Fourier domain. For this to be the case, the signal must be at least quasi-stationary. If a signal of interest is sufficiently anharmonic or nonstationary (e.g. with period changing substantially over time), it may not be captured by this approach. Additionally, this method of maximum-likelihood based fitting assumes that the data points are independent and identically distributed, which is not strictly the case for Fourier power spectra \citep[see e.g.][]{2008MNRAS.385.1279B, 2010MNRAS.402..307V, 2015MNRAS.450.2052S}. Therefore, absolute statements about the significance or probability of a detection must be treated with caution.

\begin{figure}
\begin{center}
\includegraphics[width=8.5cm]{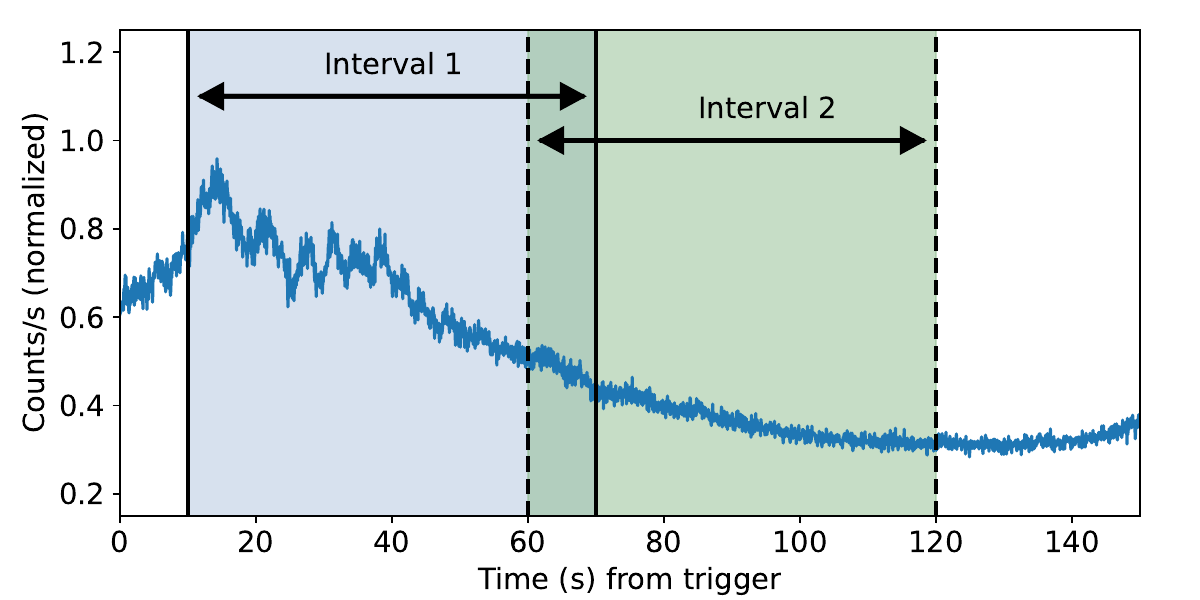}
\caption{An example of AFINO applied in an overlapping interval fashion to 15 - 25 keV GBM X-ray data from a solar flare that occurred on 2013 October 28 at 11:35 UT. Here, a QPP signal is detected in interval 1 (blue region), but not in interval 2 (green region).}
\label{windowed_afino_example}
\end{center}
\end{figure}

We apply AFINO to each event in the GBM solar flare burst catalog. It is unlikely for a short-period (P < 5s) quasi-periodic signal to be present throughout the entire or even a substantial portion of the flare evolution. Instead, we expect any QPP signature to be a transient event that may last for only a few oscillation cycles. Therefore, to maximise the chances of detecting such a signal, we apply AFINO to the solar flare timeseries using an overlapping interval approach, rather than analysing the entire flare time series at once. This is similar to a windowed Fourier transform or a dynamic power spectrum approach \citep[e.g.][]{2019ApJ...881...39H}. We first identify the beginning of the 0.064s burst mode data for each event. Then, beginning 10 s after the burst mode trigger, we analyse the flare timeseries in a sequence of 60 s intervals, stepping forward by 50s each time such that there is slight overlap between each analysed timeseries interval (see Figure \ref{windowed_afino_example}). The choice of a 60s window balances the intention to search sub-regions of the flare with the need for a sufficient number of data points in each temporal window. While other window size choices are possible, a 60 s window results in a Fourier power spectrum covering 2.5 decades in frequency space. This ensures that the model comparison method is not biased towards a null result due to a lack of data. The 50 s step interval is used to ensure that a potential QPP signal is not missed due to edge effects. In principle, a smaller step size could be used to attempt to capture frequency evolution of an oscillation. However, this would only be effective if the oscillation duration were comparable to or longer than the chosen temporal window size, which is unlikely to be the case here.
Based on these temporal window choices, the first timeseries interval runs from 10s after the burst mode trigger to 70s after the trigger, while the second interval runs from 60s after the trigger to 120s after the trigger. This process is repeated until the cadence reverts to the nominal value of 0.256s. For an individual flare the maximum number of intervals analysed is eleven, assuming 600s of burst mode data.

\section{Testing on simulated X-ray lightcurve data}
\label{simulations}

Before we search the Fermi/GBM dataset in search of short-period signals, it is important to understand whether our analysis approach is effective at detecting such signals. To test the ability of AFINO to detect a transient oscillatory signal with $P$ < 5s in a longer timeseries, we construct sets of simulated lightcurves containing an oscillation localized in time. These lightcurves are generated with a sampling time of $dt$ = 0.064s, and a total length of $t$ = 300s, mimicking the Fermi/GBM lightcurve data.

The long-term flare lightcurve evolution is represented by a two-sided Gaussian, where the width of the rising phase side of the Gaussian is narrower than the decay phase side. This approach has been previously suggested as an appropriate approximation of a flare shape \citep[e.g.][]{2019MNRAS.482.5553J, 2019ApJS..244...44B}. To this we add Poisson noise of varying strengths, and a localized oscillation. The oscillation is defined by a sine wave multiplied by a Gaussian envelope, which is added to the baseline signal at a randomized time $t_p$. The width of the envelope is defined as 3 times the oscillation period, such that the same number of oscillation cycles are always present in the data. This synthesized signal may be written as,

\begin{equation}
F(t) = F(t)_{osc} + F(t)_{bkg} + F(t)_{noise}
\end{equation}

$F_{bkg}$ is defined as,

\begin{equation}
F(t)_{bkg} = \begin{cases}
    A_0 \exp \left(-{\frac{(t - \mu)^2}{2\sigma_1^2}}\right), & \text{if $t$ $<$ $\mu$} \\
    A_0 \exp \left(-{\frac{(t - \mu)^2}{2\sigma_2^2}}\right), & \text{if $t$ $>$ $\mu$} \\
\end{cases}
\end{equation}

where $A_0$ is the amplitude of the long-term flare background, $\mu$ is the time of the peak of this background component, $\sigma_1$ is the left-side width of this component, and $\sigma_2$ is the right-side width of this component.

The oscillation component of the signal, $F_{osc}$, may be written as,

\begin{equation}
F(t)_{osc} = A_{osc} \sin(2 \pi f t) \times \exp \left(-{\frac{(t - \mu_{osc})^2}{2\sigma_{osc}^2}} \right)
\end{equation}

where $A_{osc}$ is the amplitude of the oscillation, $f$ is the frequency of oscillation, $\mu$ is the centroid location of that oscillation in the timeseries, and $\sigma_{osc}$ is the width of the Gaussian envelope for the signal, defined as three times the oscillation period, or $\sigma_{osc}$ = 3$\times$ $1/f$.

The final signal component, $F(t)_{noise}$, is drawn at each $t$ from a Poission distribution with an expectation value $\lambda$. In this work, we define the signal-to-noise ratio as the ratio between the oscillation amplitude $A_{osc}$ and the standard deviation of the Poisson noise, i.e. SNR = $A_{osc}$ / $\sqrt(\lambda)$.

\begin{figure*}
    \begin{center}
        \includegraphics[width=8.5cm]{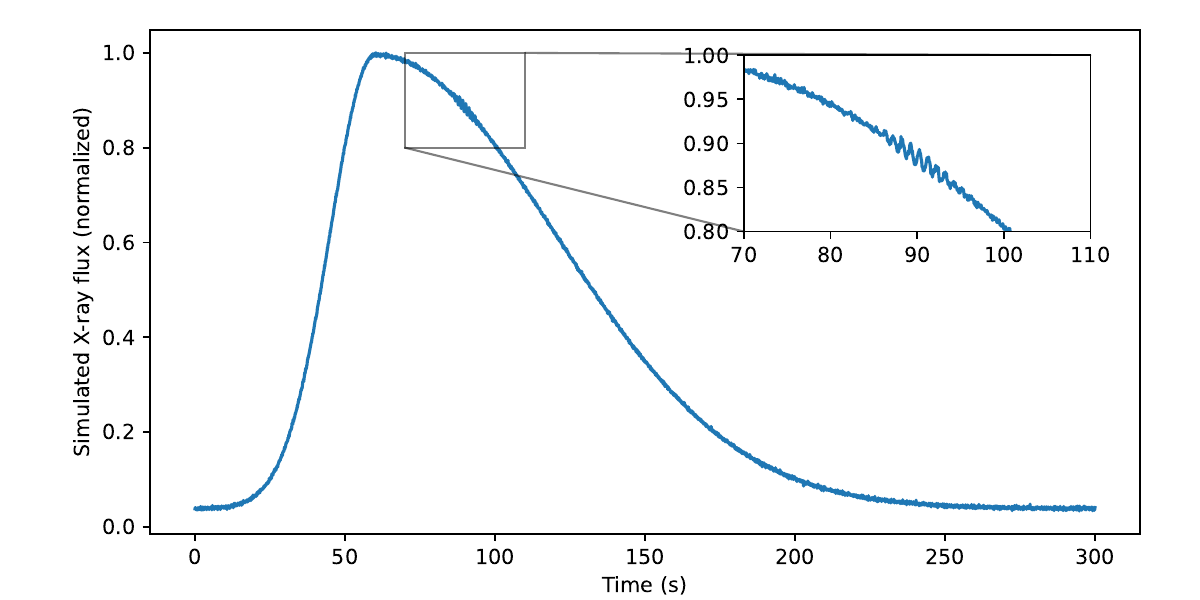}
        \includegraphics[width=8.5cm]{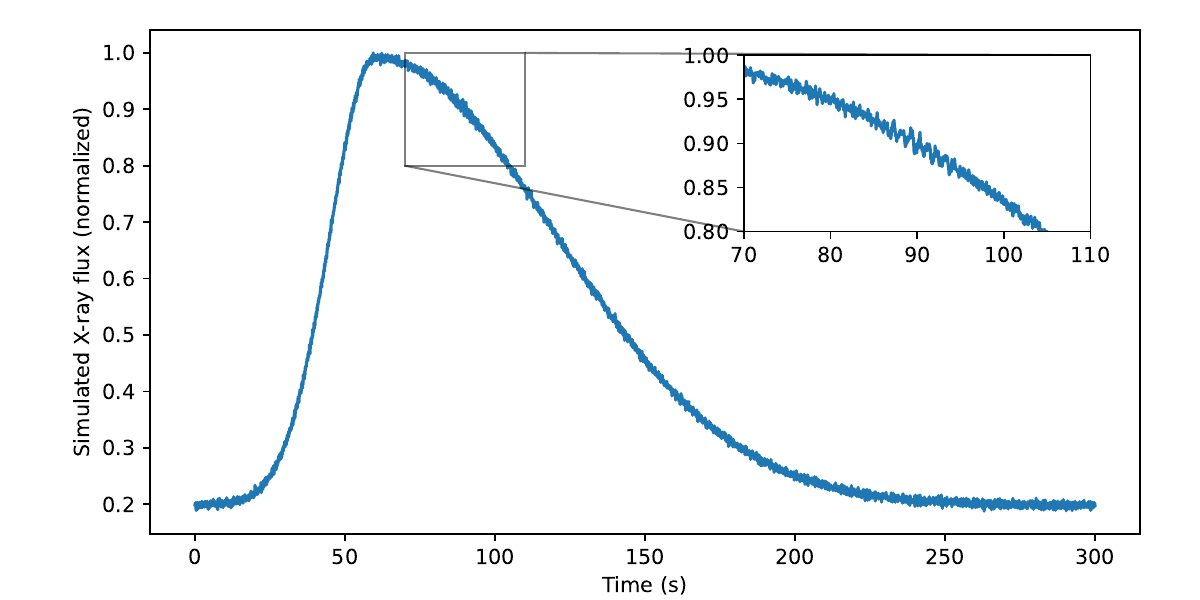}
        \caption{Examples of simulated X-ray lightcurve data, normalized by their maximum value. The data is simulated with a cadence of 0.064s (16 Hz). A two-sided Gaussian is used to represent the base flare lightcurve, to which noise is added. A transient oscillation is defined by a sine wave multiplied by a Gaussian envelope and is inserted at $\mu_{osc}$ = 90s (see inset), with a Gaussian envelope width of $\sigma$ =  3 times the oscillation period. the left panel shows a simulated lightcurve with SNR = 5. The right panel shows thee same simulated flare but with SNR = 2. In both cases, AFINO correctly identifies the oscillation with $P$ = 1 s.}
        \label{simulated_lightcurve_examples}
    \end{center}
\end{figure*}

Figure \ref{simulated_lightcurve_examples} shows two examples of AFINO applied to the simulated lightcurves. In both cases, the relevant temporal window is shown in the inset, where the transient oscillation is present. In this example, $\sigma_1$ = 15 is the width of the rise phase Gaussian background, and $\sigma_2$ = 60s is the width of the decay phase Gaussian background. The oscillation in this case is centered at $\mu_{osc}$ = 90s, which is at the center of one of the AFINO time series segments. The left panel shows the case where the signal amplitude is five times the noise amplitude. The right panel shows the case where the signal is only two times the background noise level. In both cases, the transient oscillation is successfully detected.

For our first test, we set $\mu_{osc}$ to a fixed value of 90s, so that we examine the impacts of changing the noise level and oscillation frequency in isolation. We apply the windowed AFINO method exactly as described above, and examine the results. We test noise-to-signal (1 /SNR) ratios of 0.05, 0.1, 0.2, 0.3, 0.4, 0.5, 0.6, 0.7, 0.8, 0.9, 1.0, 1.25, 1.5, 1.75, and 2.0. We test each of these SNRs for oscillation frequencies of 0.25 Hz, 0.5 Hz, 1.0 Hz, 1.5 Hz, 2.0 Hz, 2.5 Hz, and 3.0 Hz (i.e. 0.3 -- 4s period). 

\begin{figure*}
    \centering
    \includegraphics[width=8.5cm]{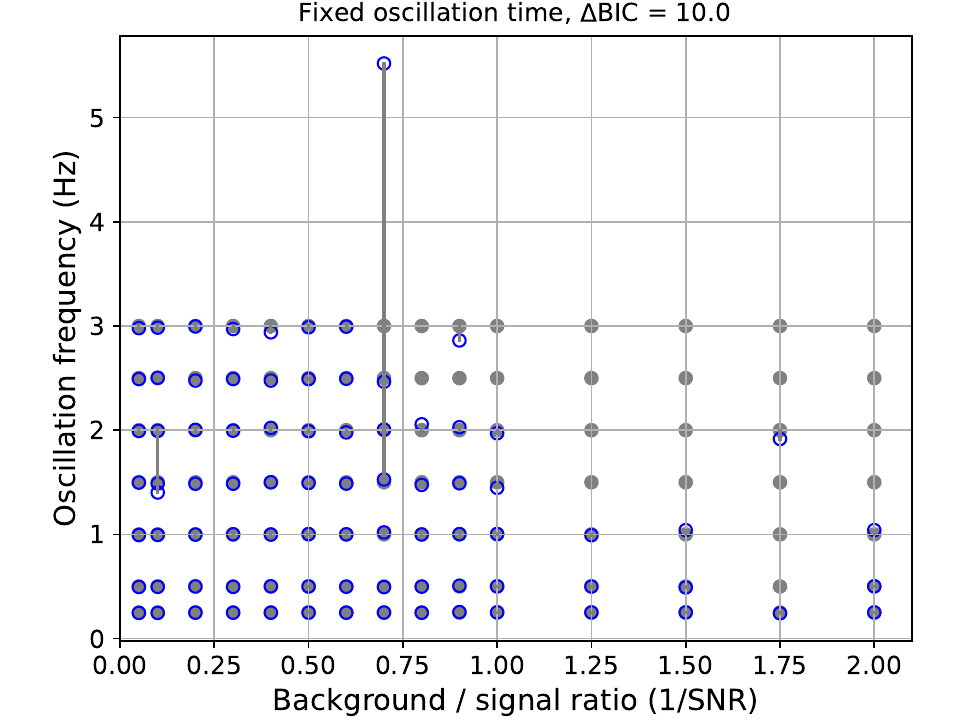}
    \includegraphics[width=8.5cm]{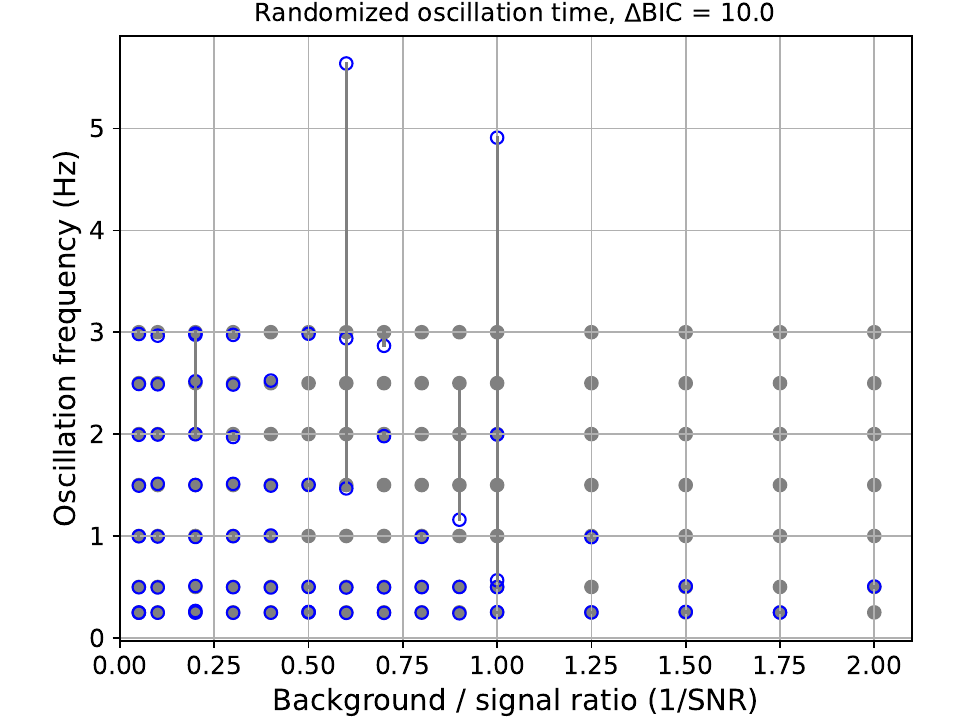}
    \includegraphics[width=8.5cm]{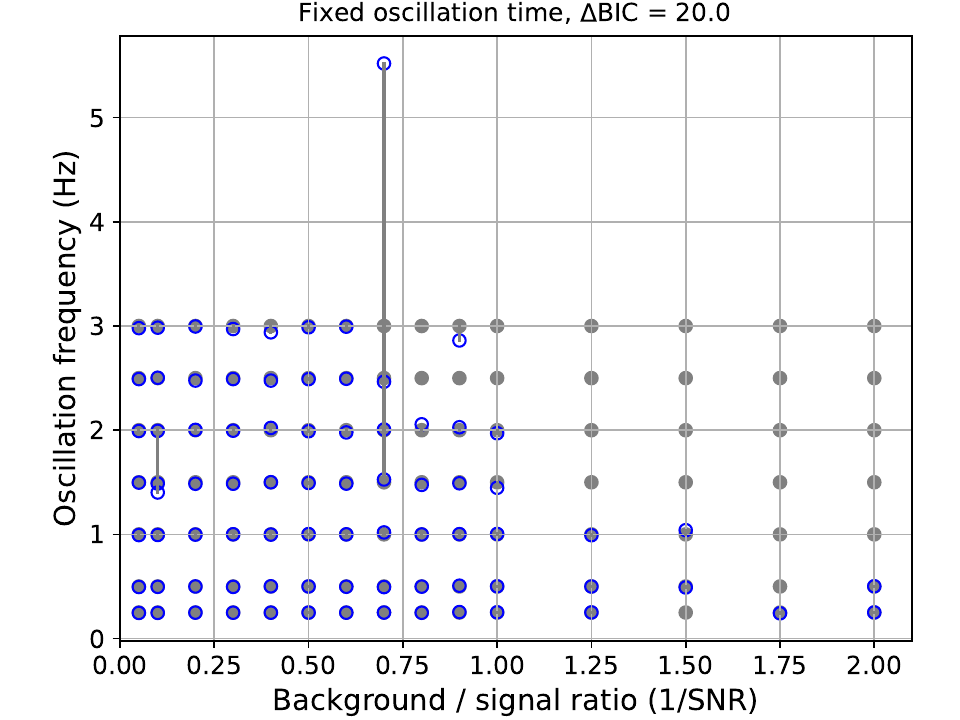}
    \includegraphics[width=8.5cm]{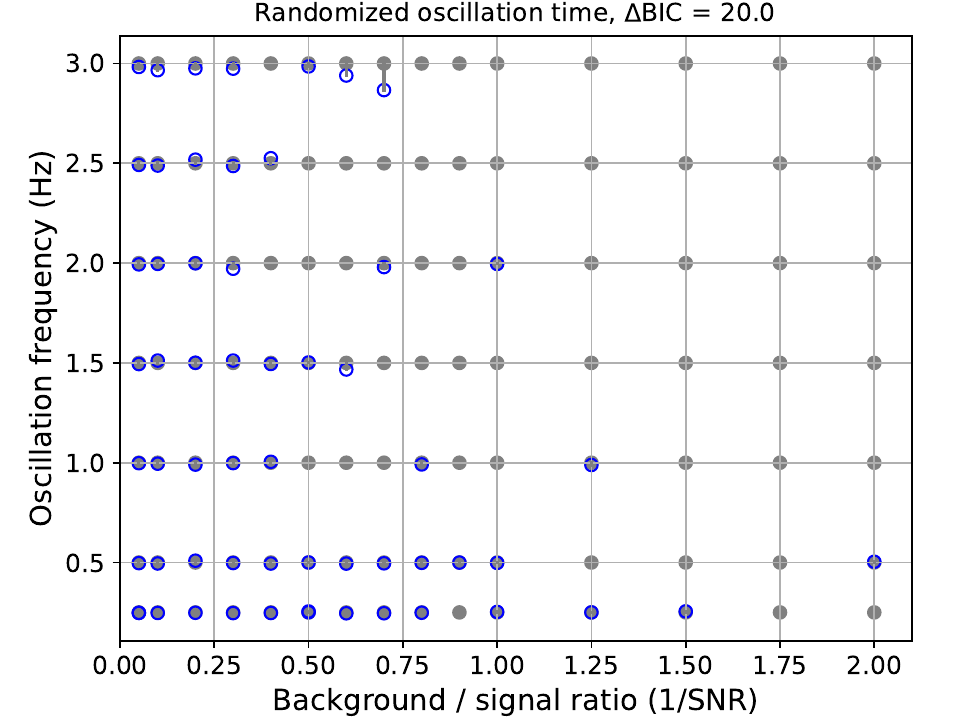}
    \caption{Detections of oscillations in simulated X-ray lightcurves.  left panels: detected frequency (blue circles) vs original simulated oscillation frequency (grey circles) as a function of the signal to noise ratio. In these panels, the transient oscillation is inserted at a fixed time in the flare timeseries. Right panels: same as left panels, except that the timing of the transient oscillation is randomized. The top row shows the results when the detection threshold is set to $\Delta$BIC > 10. The bottom row shows the results of the same experiment when the detection threshold is set to $\Delta$BIC > 20.}
    \label{simulated_results_fixed_oscillation}
\end{figure*}

The top left panel of Figure \ref{simulated_results_fixed_oscillation} illustrates the results of this test using a standard detection threshold of $\Delta$BIC > 10 \citep{2016ApJ...833..284I, 2020ApJ...895...50H}. The grey circles indicate each combination of oscillation frequency and background noise level tested. The blue circles indicate instances where the oscillation was successfully detected by AFINO using the running window technique, and the actual frequency detected.

 For pulsations of 0.25 Hz and 0.5 Hz (P = 4s, P = 2s), we see that the transient oscillation is detected by AFINO even when the noise level exceeds the signal level (SNR < 1). For pulsation frequencies of 1 Hz, the transient oscillation is also mostly detected while SNR > 1, with a few detections missed. As frequency increases, the signal becomes harder to detect due to it's shorter total duration in the lightcurve (since the Gaussian envelope is a fixed function of the oscillation frequency). At 3 Hz, the signal can only be reliably detected once the 1/SNR value reaches 0.6. There are also two noticeably inaccurate detections during this test, where the detected period is different from the true period. 

Having established the effect of the signal-to-noise ratio on detecting a period at a fixed $\mu_{osc}$, we next attempt to detect a localized oscillation in cases where $\mu_{osc}$ is randomized. This represents a more realistic scenario that will be encountered in real solar X-ray data. We test the same set of frequencies and noise levels as before. The results are shown in the right hand panels of Figure \ref{simulated_results_fixed_oscillation}. 

We see that, as before, signals at lower frequencies are easier to detect when the signal to noise ratio is low. At $f$ = 0.25 Hz and $f$ = 0.5 Hz, the signal is still detectable at times when SNR < 1. Generally speaking, signals with $f$ > 1 Hz can only be detected when SNR > 1, and only reliably detected when SNR > 2 (or 1/SNR < 0.5).

The bottom row of Figure \ref{simulated_results_fixed_oscillation} repeats both experiments with a stricter detection threshold of $\Delta$BIC > 20. In this scenario, we see that inaccurate detections of the input signal -- where the reported frequency is incorrect -- are almost eliminated. Meanwhile the correct signals are still detected at similar SNR ratios as before.

These simulations show that AFINO -- when applied using overlapping analysis windows -- is able to detect short-period <5s pulsations in flare-like lightcurves for a range of background noise levels and oscillation frequencies, even in the case where the oscillation is transient and only present for a short time. This was demonstrated both for the case of an oscillation at a fixed point in the flare decay phase, and also for oscillations where the temporal location was randomized in time.

\subsection{Testing the false alarm rate and the true positive rate}
\label{false_alarm}

In real Fermi/GBM solar X-ray data, the true number of quasi-periodic signals is unknown. Since we are searching a large number of timeseries, and applying a stepping interval approach to each one, the total number of signals analysed by AFINO is large, over 15,000 in each energy band. It also follows that, even if a QPP is present in a flare, many of the individual flare time series segments analysed will not contain an oscillatory signal. Therefore, it is crucial to understand to the extent possible the rate of false alarms in the detection method, in other words the rate of type 1 errors. When analysing so many individual time series, if the true occurrence rate is low even a modest false alarm rate may dominate the results, a phenomenon known as the base rate fallacy.

We can test the false alarm rate of our approach by applying AFINO to a large number of simulated lightcurves that by design do not contain an oscillatory signature (see \citet{2019A&A...629A.147B} for a similar concept). We construct 1000 flare-like lightcurves of the form described in Section \ref{simulations} above, with some key differences. First, no transient oscillation is inserted into the signal. Second, the width parameters of the two-sided Gaussian that determines the background flare shape are allowed to vary, creating variation in the simulated signals. Finally, the background noise level is randomized for each simulated signal. 

Complementary to this, we can also explore the impact of changing the detection threshold on the true positive detection rate. We achieve this in the same manner as the false alarm testing. We construct 200 flare-like signals with randomized two-sided Gaussian background profiles, but this time containing a randomized, local pulsation as described in Section \ref{simulations} above and added noise at a fixed signal-to-noise ratio. We repeat this for a number of SNR values, i.e. each SNR value is associated with 200 flare-like signals. We then apply AFINO to these synthetic signals and evaluate the number of times the correct oscillatory signal was identified.

\begin{figure}
\begin{center}
\includegraphics[width=8.5cm]{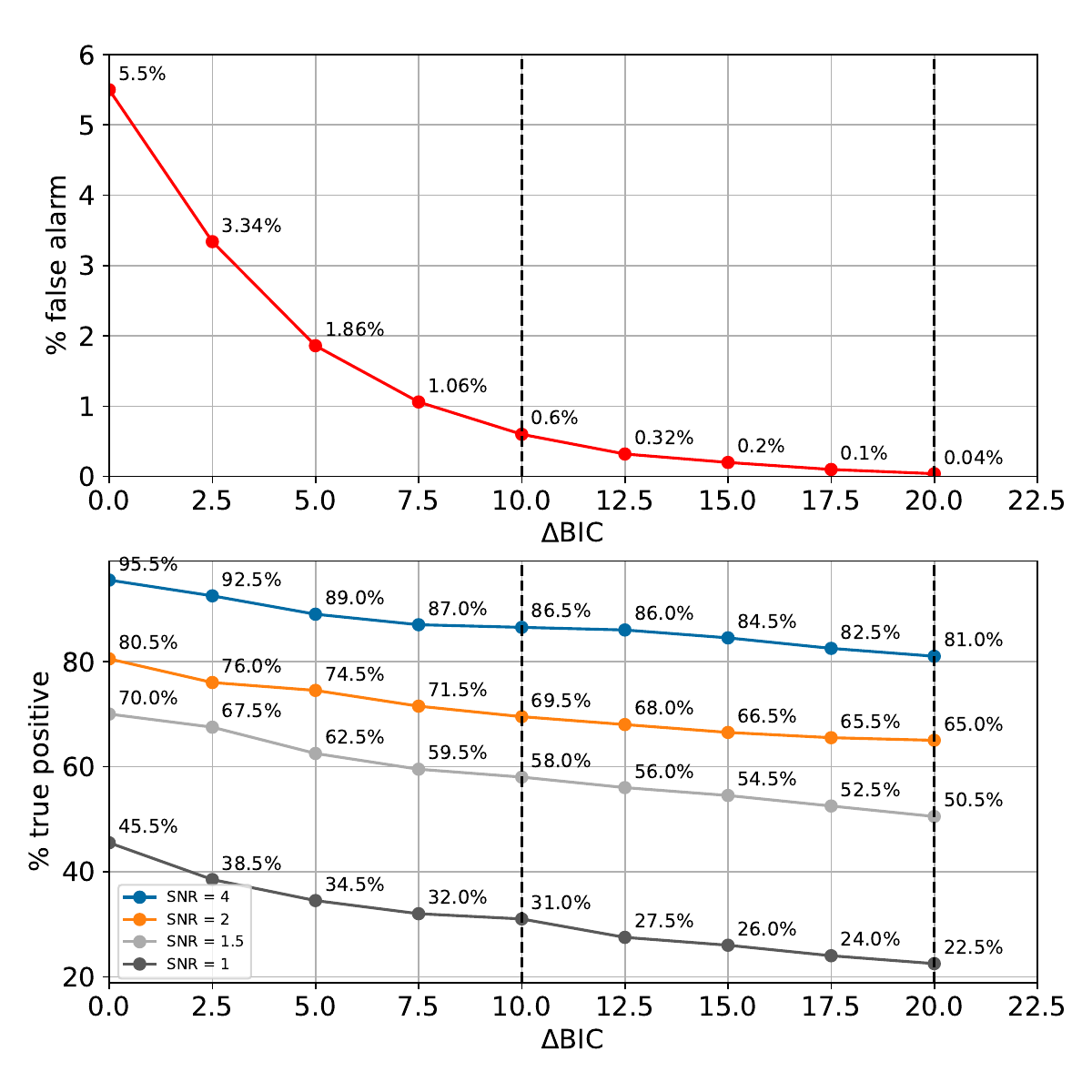}
\caption{Top: the false alarm rate of AFINO as a function of the detection threshold parameter $\Delta$BIC. This is estimated by applying AFINO to a suite of simulated X-ray lightcurves that do not contain an oscillation. Bottom: The true positive detection rate of AFINO as a function of $\Delta$BIC, for three different signal-to-noise (SNR) ratios. This is estimated by applying AFINO to a suite of simulated X-ray lightcurves containing oscillations, and measuring the number of correct identifications.}
\label{false_alarm_vs_bic}
\end{center}
\end{figure}

We apply AFINO to each simulated timeseries -- both the oscillation-free signals and those containing a QPP -- using the stepping window technique as before. Thus, each individual timeseries is broken up into five segments. In the case of the oscillation-free signals, AFINO is applied a total of 5,000 times, allowing us to investigate low false-alarm rates. By testing different $\Delta$BIC thresholds for identifying a significant QPP event, we can evaluate the false alarm rate and the true positive rate (see Figure \ref{false_alarm_vs_bic}). 

Firstly, regarding the false positive rate we see that, when a threshold of $\Delta$BIC $>$ 0 is used, we obtain a large false alarm rate of $>$ 5.5\%. When a more typical detection threshold of $\Delta$BIC > 10 is used, this resulted in 30 false alarms, a rate of 0.6\%. Although low, in the analysis of real Fermi/GBM data where we expect the number of true positives to be low we must be concerned about false alarms being a significant or dominant contribution to the total number of detections. We are analysing 1460 individual solar flares, each of which is broken up into a maximum of eleven individual time series segments. AFINO is being applied a total of 15,358 times in each energy band studied. With a false alarm rate of 0.6\%, we could therefore expect $\sim$ 90 false alarms in each energy band if the GBM data behaves similarly to the simulated data. If the detection threshold is raised to a conservative value of $\Delta$BIC $>$ 20, we see that the false alarm rate is much reduced, to $\sim$ 0.04\%. 

We can evaluate the true positive rate in a similar fashion (see Figure \ref{false_alarm_vs_bic}, bottom panel). For a true positive detection, we require that the period identified by AFINO must match the true underlying period to within a tolerance of 25\%. We see that the overall true positive detection rate is strongly dependent on the SNR as expected. However, for a given SNR, the reduction in true positive detections as $\Delta$BIC increases is modest compared to the behaviour of the false alarm rate. For signals with SNR = 2 for example, at $\Delta$BIC > 0 the true positive rate is 80.5\%, while at $\Delta$BIC > 10 it drops to 69.5\%, and at $\Delta$BIC > 20 it is reduced to 65.0\%.

How this affects the statistical results of a QPP search strongly depends on the true base occurrence rate of QPP signals in solar flares, which is unknown and has widely varying estimates \citep{2021SSRv..217...66Z}. However, these simulations show that adopting a more conservative $\Delta$BIC value can strongly reduce the false alarm rate with only a modest cost to the true positive detection rate. Therefore, in the analysis of the real GBM X-ray data, we focus on detection of QPPs using strict thresholds of $\Delta$BIC > 15 or $\Delta$BIC > 20.

\section{Analysis of Fermi/GBM X-ray flare trigger data}
\label{gbm_data_analysis}

\subsection{Statistical results}
\label{statistical_results}

As described in Section \ref{data_and_methods}, we apply AFINO to 1460 flares observed by Fermi/GBM in trigger mode. For each flare, we split the timeseries up into a number of overlapping intervals of 60s length, beginning 10s after the trigger time, and analyse each interval independently. This is done in three energy channels; 4 -- 15 keV, 15 -- 25 keV, and 25 -- 50 keV. For a flare with 600s of burst mode data, this results in 11 time series analysis segments. In each energy range, the total number of independent time series intervals studied with AFINO is 15358.

In total, using a strict detection threshold of $\Delta$BIC > 20, we find strong evidence of rapid pulsations for 13 flares in the 4 -- 15 keV energy band, 8 flares in the 15 -- 25 keV band, and 5 flares in the 25 -- 50 keV band. 

Considering our estimated false alarm rate of 0.04\% based on simulated lightcurves with a threshold of $\Delta$BIC = 20, the expected number of false alarms in each energy band is $\sim$ 6. In the 4 -- 15 keV energy band, we do identify more QPP-like signals than expected from the false alarm rate, but the difference is only a factor of 2. For the other energy bands, the detection rate is inline with that expected from false alarms. Thus, while we may have identified some real QPP signals, we do not find strong evidence that such signals are widespread with periods in the $<$ 5s regime using this approach.

If we adopt a less stringent threshold of $\Delta$BIC = 15, we find instead 39 events in the 4 -- 15 keV band, 17 events in the 15 -- 25 keV band, and 11 events in the 25 -- 50 keV band. Comparing again with the estimated false alarm rate -- this time of 0.2\% -- the expected number of false alarms is $\sim$ 30 in each energy band. Thus, again our results are inline with false alarm rate expectations, and we have not found strong evidence of a large underlying short-period QPP population.

\begin{figure*}
\begin{center}
\includegraphics[width=8.5cm]{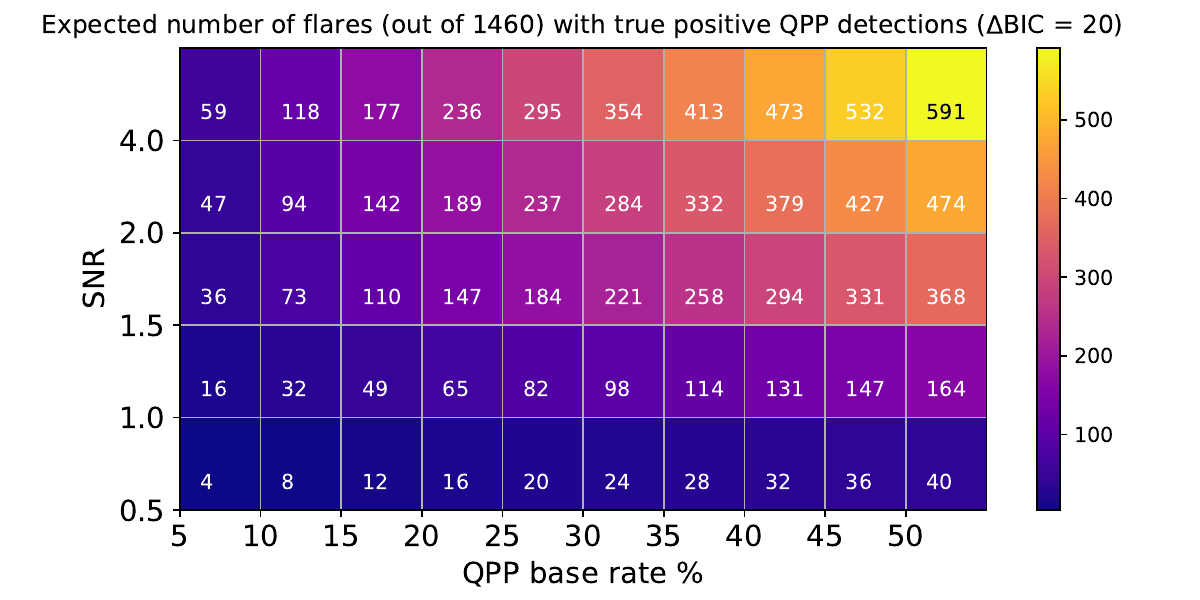}
\includegraphics[width=8.5cm]{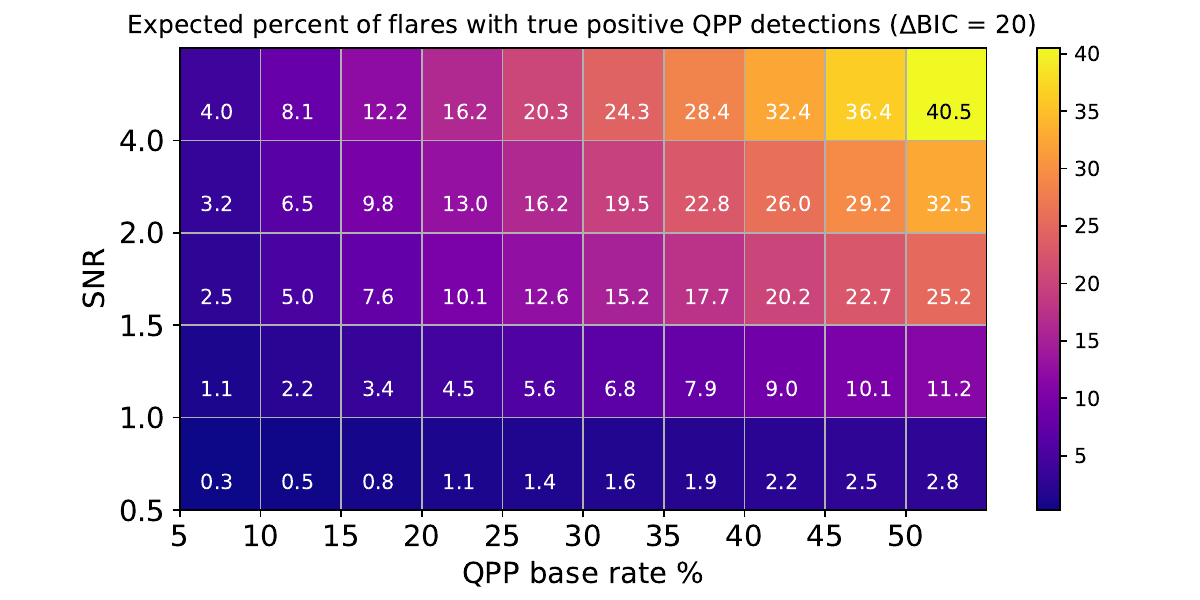}
\includegraphics[width=8.5cm]{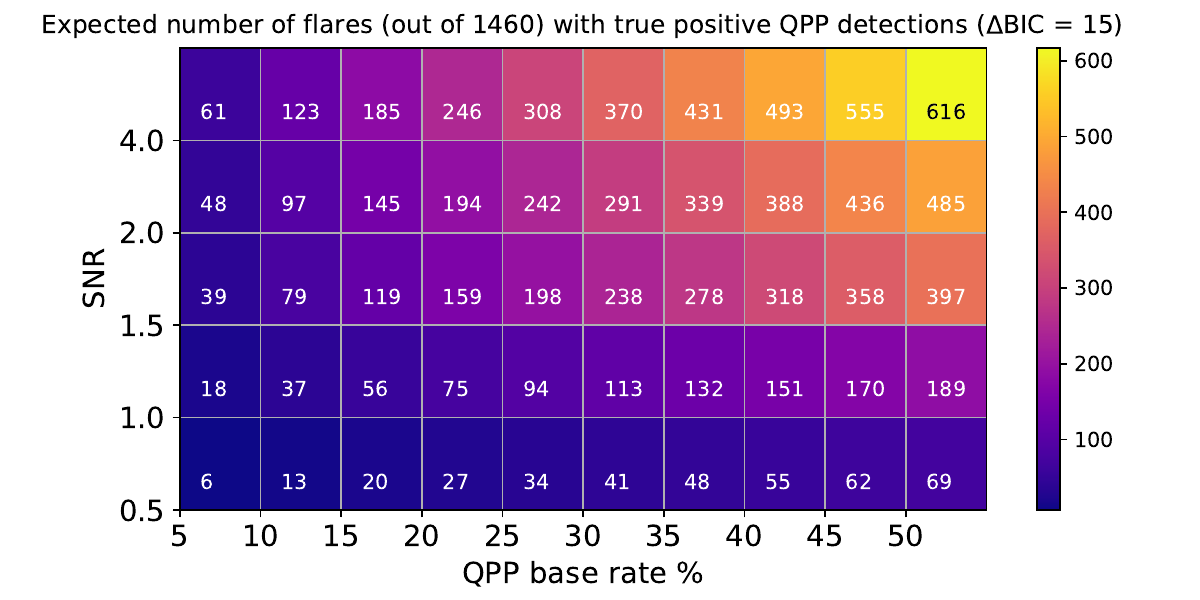}
\includegraphics[width=8.5cm]{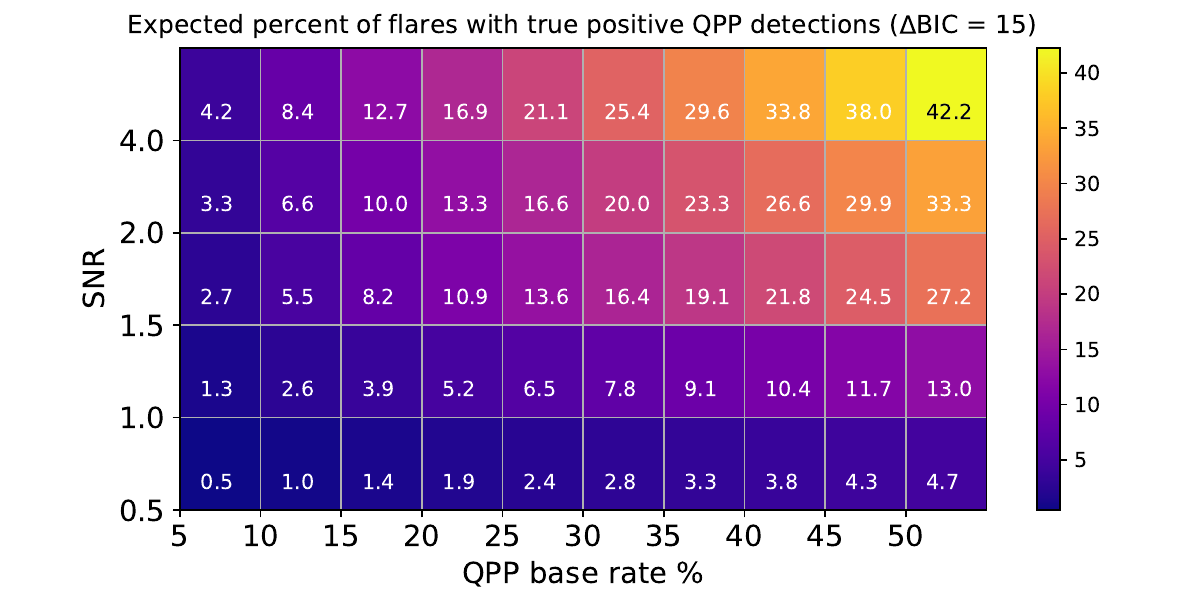}
\caption{The expected detection rate of short-period QPP in flares as a function of the QPP base rate and the QPP signal-to-noise ratio, based on synthetic lightcurve analysis (see Section \ref{false_alarm}). Top left: The expected number of flares, out of 1460, featuring a QPP detection, as a function of SNR and base rate, using a detection threshold of $\Delta$BIC > 20. Top right: The expected percentage of flares featuring a QPP detection as a function of SNR and base rate. Bottom row: Same as top row, but for a less stringent detection threshold of $\Delta$BIC > 15.}
\label{base_rate_plot}
\end{center}
\end{figure*}

As our application of AFINO to simulated flare lightcurves showed (see Figure \ref{false_alarm_vs_bic}), the expected true positive rate of QPP detections is strongly dependent on the mean signal-to-noise ratio of the QPP signals. It was also shown in \citet{2019ApJS..244...44B} that, while AFINO had a low false alarm rate, it was also more conservative regarding real detections than other methods. Therefore, it is likely that there are a number of real QPP signals present in this flare dataset that we were unable to positively identify.

We can explore this further in order to put constraints on the possible QPP base rates in the GBM data as well as the likely signal-to-noise ratios of those signals in the GBM burst mode dataset. We achieve this by using the true positive detection rates estimated from simulations of QPP-like signals in synthetic flares described in Section \ref{false_alarm} and shown in Figure \ref{false_alarm_vs_bic}. By assuming different base occurrence rates of QPPs in GBM flare data, and using our estimates of true positive detection rates as a function of signal-to-noise ratio, we can estimate the percentage of detections we would expect from the GBM flare dataset, as well as the absolute number of detections. This is shown in Figure \ref{base_rate_plot}.

The top row of Figure \ref{base_rate_plot} shows the expected number (left panel) of true positive detections  -- out of 1460 solar flares -- as a function of the assumed QPP base rate and the SNR of those QPP signals. The right panel expresses the same result as a percentage. These values are calculated assuming that an individual QPP signal is present in only a single flare time series segment. For the top row, this is estimated using a $\Delta$BIC > 20 threshold. The bottom row provides the same estimates for a less stringent threshold of $\Delta$BIC > 15. 

Comparing the number of QPP events we have detected in the GBM data with Figure \ref{base_rate_plot} clearly indicates that observationally we have some combination of a low SNR and low base rate of QPPs. Either the true number of short-period QPP signals present in the GBM dataset is low, or the typical SNR of signals is sufficiently low (i.e. SNR < 1) that detection is very rare. These results show that short-period QPPs with strong signal-to-noise ratios would have been detected in large numbers, even for modest base rates of 10--20\%. For a base rate of 20\% percent and an SNR of 2.0 for example, 189 true positive detections would be expected using a $\Delta$BIC > 20 threshold, and 194 using a $\Delta$BIC > 15 threshold. 

In summary, our analysis of 1460 solar flares observed with GBM burst mode data does not reveal strong statistical evidence for a large number of short-period QPPs. We surmise that either the true occurrence rate of such signals is low, or that for the burst mode data they are usually present with signal-to-noise ratios of less than 1 and are therefore challenging to detect and distinguish from false alarms.

\subsection{Case studies of individual events}
\label{case_studies}

Although our statistical study of Fermi/GBM flare data did not find evidence for a large underlying population of QPPs with periods in the $<$ 5s regime, we have nevertheless identified several interesting QPP events that we are confident are real solar signals, and not false alarms. Here, we explore a selection of these events in more detail, although we do not assign a strict statistical significance to these events. Each flare presented below showed strong evidence of QPPs in at least one GBM energy channel (meeting at least a threshold of $\Delta$BIC > 15), and additionally shows visually compelling evidence of QPPs. We identify each event using its GBM catalog burst number, which is based on the date and time (expressed as fraction of day) of the observation.  For example, BN110616429 corresponds to 2011-06-16 at 10:17:45 UT.

\subsubsection{BN110616429}

This burst was associated with a C7.1-class solar flare that occurred on 2011-06-16, beginning at 10:13 UT according to the GOES flare catalog and continuing until 10:29 UT. The Fermi/GBM trigger was at 10:17:46 UT. This was one of a series of flares occurring from AR 11237, as noted by \citet{Panesar_2015}.

\begin{figure}
\begin{center}
    \includegraphics[width=8.5cm]{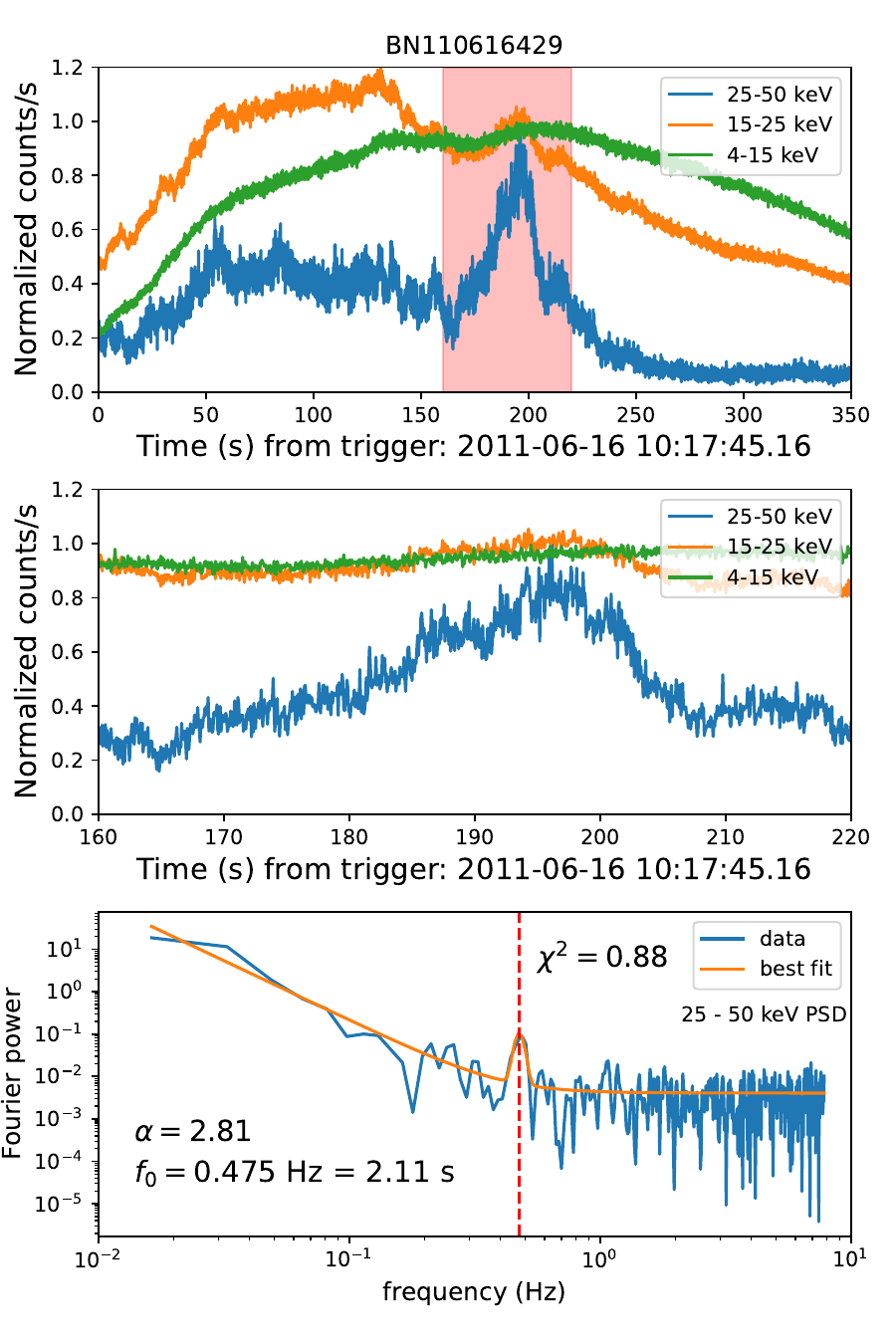}
    \caption{Summary of the BN110616429 solar flare burst. Top: Normalized count rates in the 4-15 keV (green), 15-25 keV (orange), and 25-50 keV (blue) energy ranges at 0.064s cadence. The time series segment where a significant QPP was detected is shaded in red. Center: Zoomed-in lightcurve showing only the time series segment where the QPP was detected, i.e. the red shaded area from the previous panel. Bottom: The Fourier power spectral density (PSD) of the flare time series segment (blue), with the best fit of the QPP-like AFINO model overplotted (orange). Here, a period of $P$ = 2.1s was detected in the 25 -- 50 keV energy band. In this plot and all subsequent summary plots, the absolute y-values of the GBM lightcurves have been adjusted by adding or subtracting constant factors as needed for visual clarity.} 
    \label{summary_bn110616429}
\end{center}
\end{figure}

Figure \ref{summary_bn110616429} shows the Fermi/GBM lightcurves of this event in the 4-15 keV, 15-25 keV, and 25-50 keV energy ranges. In the top panel, the red shaded area highlights the portion of the flare timeseries where the QPP was detected. The center panel shows a zoomed in timeseries of the red shaded area from the top panel. The bottom panel of Figure \ref{summary_bn110616429} shows the result of the AFINO power spectrum model fitting procedure. For this flare, a period of $P$ = 2.1s was detected in the 25 -- 50 keV energy band.

\subsubsection{BN120510849}

This burst trigger was associated with a solar flare from 2012-05-10. The flare was of GOES class M1.7 and began at 20:20 UT, continuing until 20:30 UT. The flare was sourced from AR 11476. 

\begin{figure}
\begin{center}
    \includegraphics[width=8.5cm]{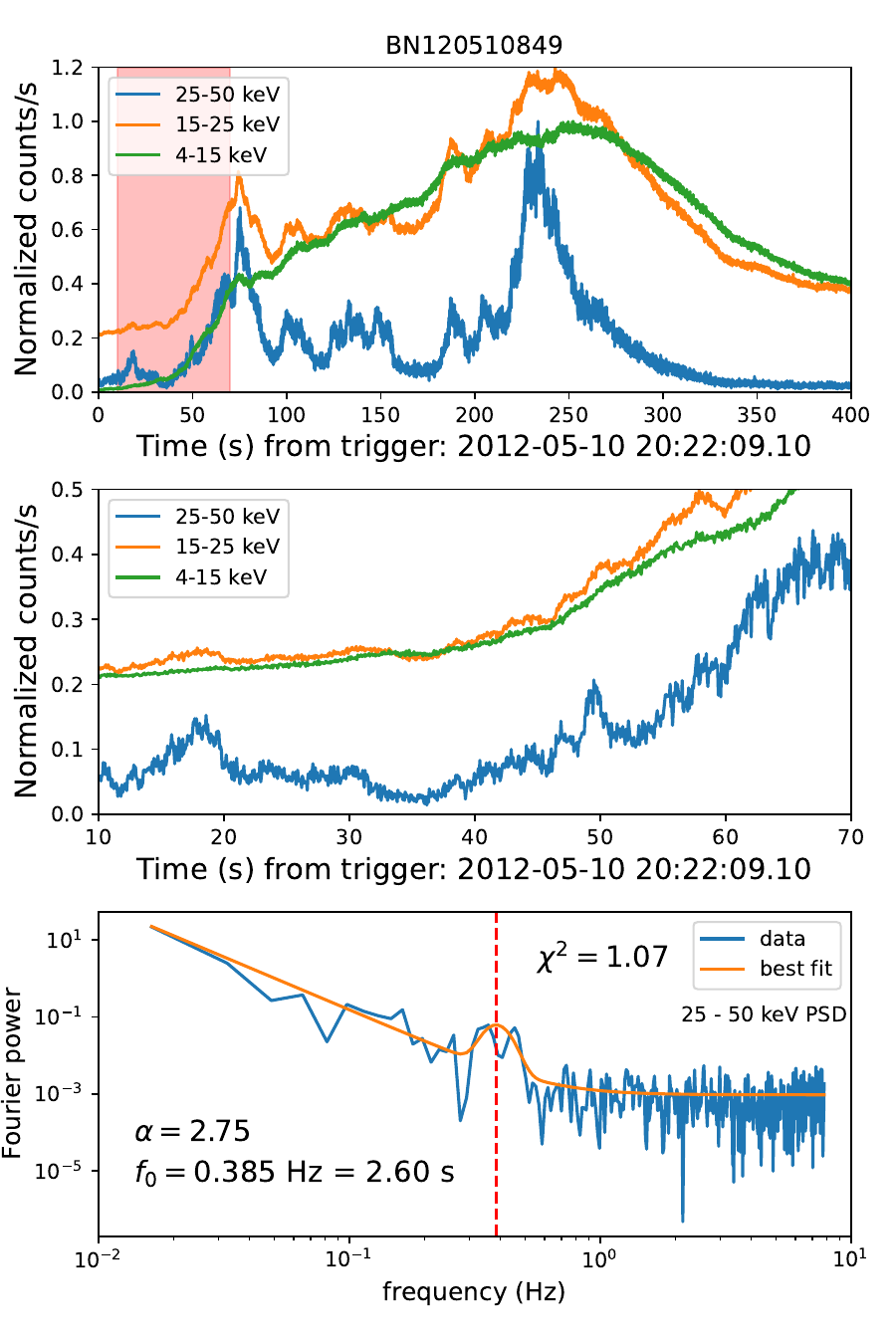}
    \caption{Summary of the BN120510849 solar flare burst. Top: Normalized count rates in the 4-15 keV (green), 15-25 keV (orange), and 25-50 keV (blue) energy ranges at 0.064s cadence. The time series segment where a significant QPP was detected is shaded in red. Center: Zoomed-in lightcurve showing only the time series segment where the QPP was detected, i.e. the red shaded area from the previous panel. Bottom: The Fourier power spectral density (PSD) of the flare time series segment (blue), with the best fit of the QPP-like AFINO model overplotted (orange). Here, a period of $P$ = 2.6s was detected in the 25 -- 50 keV energy band.} 
    \label{summary_bn120510849}
\end{center}
\end{figure}

Figure \ref{summary_bn120510849} shows the Fermi/GBM lightcurves of this event and the best-fit model to the Fourier power spectrum in the bottom panel. The periodic structure was detected in the first timeseries segment of this flare, as shown by the red highlighted area. A periodic structure with $P$ = 2.6s was detected by AFINO in the 25-50 keV band. Visually, a corresponding set of pulsations can be clearly observed between $t$ = 40s and $t$ = 50s. 

\subsubsection{BN131028192}

The GBM trigger BN131028192 is associated with a strong solar flare of GOES class M5.1, originating from AR 11875 on 2013-10-28. The Fermi trigger time for this event was 04:35:51 UT, while the GOES catalog lists the flare beginning at 04:32 UT and ending at 04:46 UT. This active region was also associated with a number of other powerful solar flares, including several others on this day, some of which were associated with longer period QPPs, on the order of 10's of seconds \citep[e.g.][]{2016ApJ...827L..30H}. 

\begin{figure}
\begin{center}
    \includegraphics[width=8.5cm]{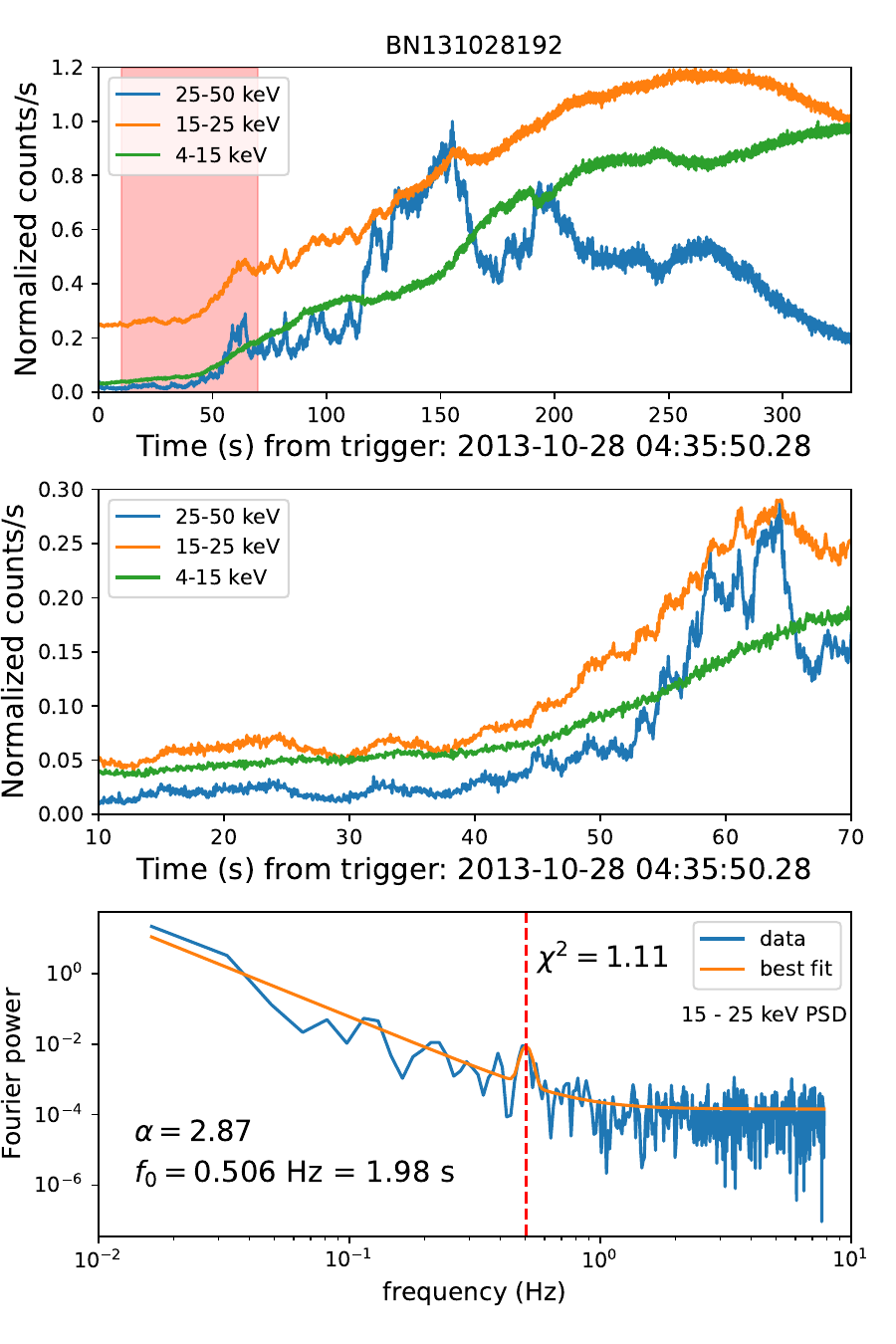}
    \caption{Summary of the BN131028192 solar flare burst. Top: Normalized count rates in the 4-15 keV (green), 15-25 keV (orange), and 25-50 keV (blue) energy ranges at 0.064s cadence. The time series segment where a significant QPP was detected is shaded in red. Center: Zoomed-in lightcurve showing only the time series segment where the QPP was detected, i.e. the red shaded area from the previous panel. Bottom: The Fourier power spectral density (PSD) of the flare time series segment (blue), with the best fit of the QPP-like AFINO model overplotted (orange). A period of $P$ = 3.7s was detected in the 15 -- 25 keV energy band.} 
    \label{summary_bn131028192}
\end{center}
\end{figure}

Figure \ref{summary_bn131028192} shows the Fermi/GBM lightcurves of the event and the best-fit AFINO result. In this case, AFINO detects a significant oscillation with $P$ = 2s in the 15 -- 25 keV energy band, during the first flare time series segment. Inspection of this segment via the center panel of Figure \ref{summary_bn131028192} shows the presence of such an oscillation in the 15 -- 25 keV lightcurve (orange curve) between approximately $t$ = 45s and $t$ = 65s after the burst trigger. Similar features are seen in the 25 -- 50 keV energy band, but are not as strongly identified by AFINO.

\begin{figure}
\begin{center}
\includegraphics[width=8.5cm]{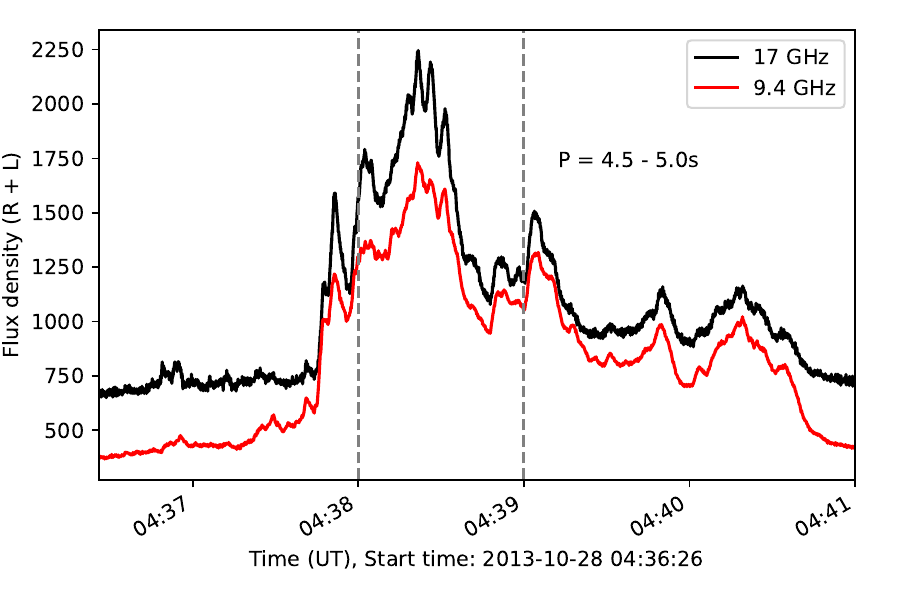}
\caption{Lightcurves of the solar flare of 2013 October 28 as observed at radio frequencies of 9.4 GHz (red) and 17 GHz (black) by the Nobeyama Radiopolarimeters. A significant period of 4.5 - 5.0s is detected by our analysis method during the interval 04:38 to 04:39 UT.}
\label{norp_20131028}
\end{center}
\end{figure}

This event was also observed by the Nobeyama Radioheliograph (NoRH) and Nobeyama Radiopolarimeters (NoRP). Both of these instruments observe in the GHz range and can provide observations with 0.1s time resolution during certain events. Figure \ref{norp_20131028} shows the 17 GHz and 9.4 GHz time series data from NoRP, showing clear bursty structure on a range of timescales. Applying AFINO to this 0.1s cadence radio data in 60s overlapping intervals reveals the presence of a period of P $\approx$ 4.5 - 5.0s betwen 04:38 and 04:39 UT. This does not temporally overlap with the period found in the GBM data, which was in the interval 04:36 UT to 04:37 UT. This may be due to the fact that the NoRP 0.1s mode data did not begin until around 04:36:30 UT. Nevertheless, the combined X-ray and radio data reveals an interesting, multi-periodic solar flare.

\subsubsection{BN131028483}

This solar flare trigger is associated with an solar flare from 2013-10-28 with a GOES class of M1.4. In the GOES catalog, the flare began at 11:32 UT and ended at 12:39 UT. The Fermi trigger time for this event was 11:35:26 UT.

\begin{figure}
\begin{center}
    \includegraphics[width=8.5cm]{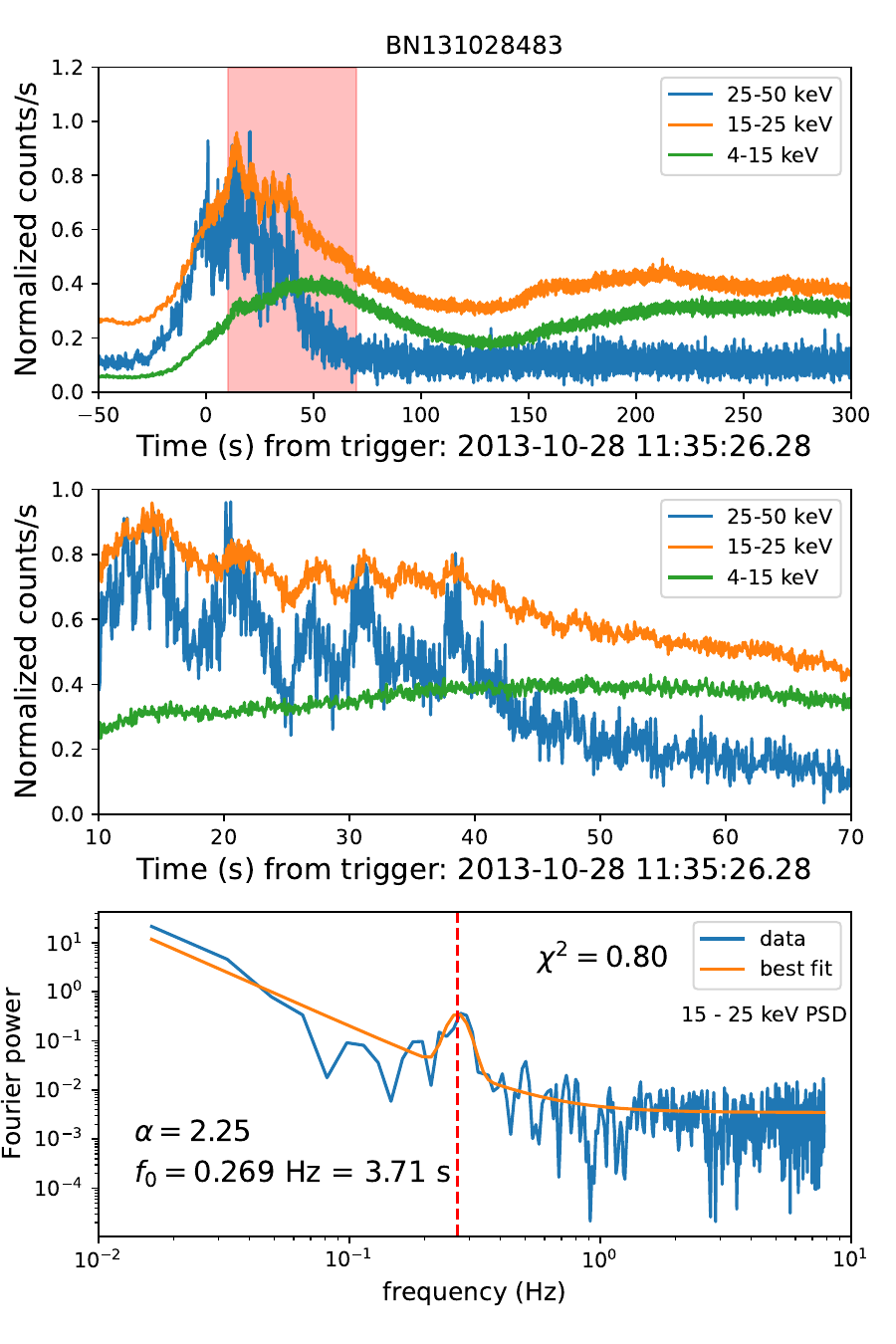}
    \caption{Summary of the BN131028483 solar flare burst. Top: Normalized count rates in the 4-15 keV (green), 15-25 keV (orange), and 25-50 keV (blue) energy ranges at 0.064s cadence. The time series segment where a significant QPP was detected is shaded in red. Center: Zoomed-in lightcurve showing only the time series segment where the QPP was detected, i.e. the red shaded area from the previous panel. Bottom: The Fourier power spectral density (PSD) of the flare time series segment (blue), with the best fit of the QPP-like AFINO model overplotted (orange). A period of $P$ = 3.7s was detected in the 15 -- 25 keV energy band.} 
    \label{summary_bn131028483}
\end{center}
\end{figure}

Figure \ref{summary_bn131028483} shows the Fermi/GBM lightcurves of this event and the best-fit AFINO model as before. A periodic structure is detected in the first timeseries segment of this flare, with $P$ = 3.7s. These pulsations are most strongly evident in the 15 - 25 keV energy band. In the center panel of Figure \ref{summary_bn131028483}, a clear sequence of pulsations can be seen from $t$ = 10s to $t$ = 50s, consistent with the AFINO detection. These pulsations can also be seen visually in the 25 -- 50 keV band.

\subsubsection{BN141020790}

This burst trigger was associated with a solar flare from 2014-10-20. The flare was of GOES class M1.4 and began at 18:55 UT, continuing until 19:04 UT. The source of the solar flare was AR 12192. The Fermi trigger time was 18:57:08 UT.

\begin{figure}
\begin{center}
    \includegraphics[width=8.5cm]{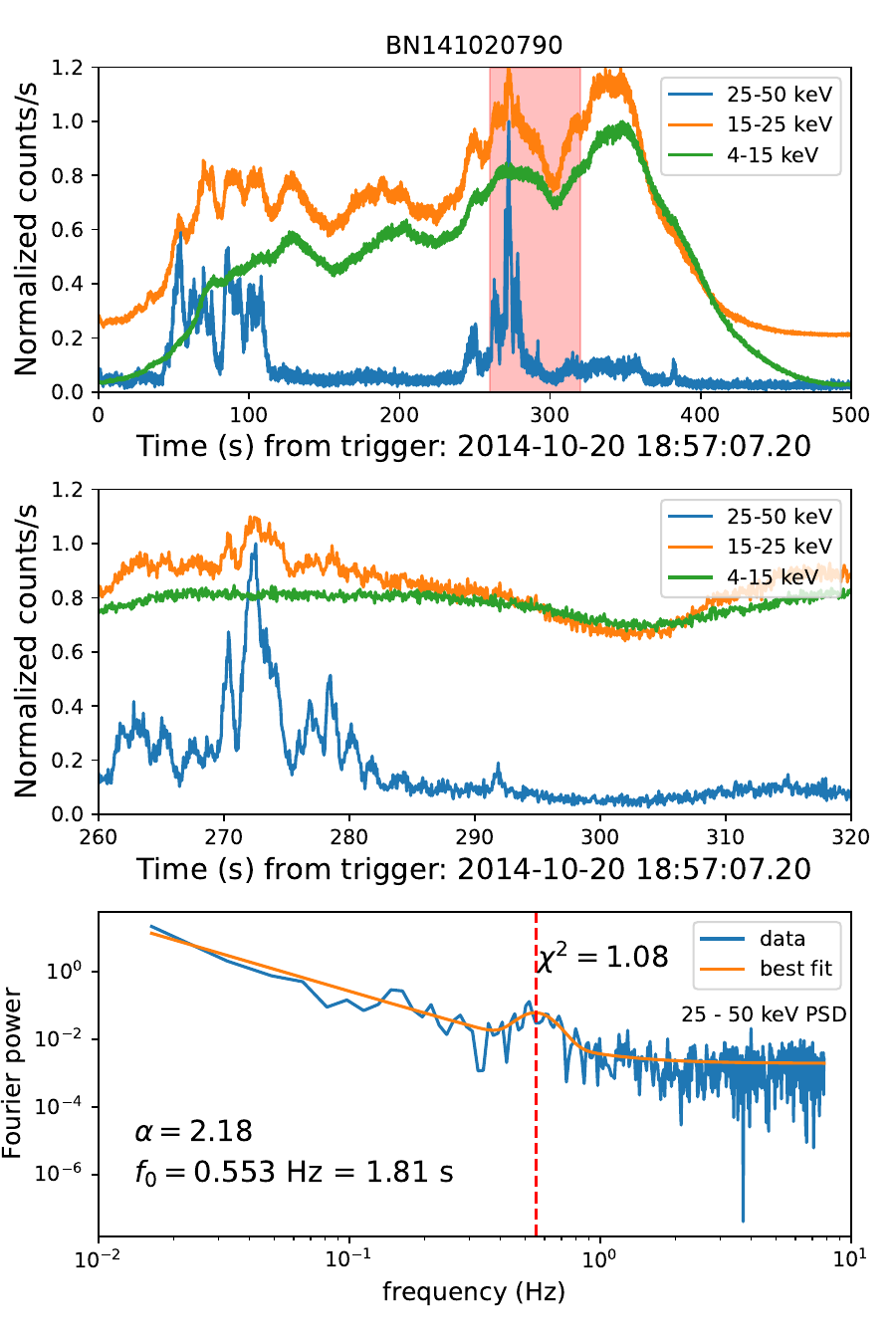}
    \caption{Summary of the BN141020790 solar flare burst. Top: Normalized count rates in the 4-15 keV (green), 15-25 keV (orange), and 25-50 keV (blue) energy ranges at 0.064s cadence. The time series segment where a significant QPP was detected is shaded in red. Center: Zoomed-in lightcurve showing only the time series segment where the QPP was detected, i.e. the red shaded area from the previous panel. Bottom: The Fourier power spectral density (PSD) of the flare time series segment (blue), with the best fit of the QPP-like AFINO model overplotted (orange). Here, a period of $P$ = 1.8s was detected in the 25 -- 50 keV energy band.} 
    \label{summary_bn141020790}
\end{center}
\end{figure}

Figure \ref{summary_bn141020790} shows the Fermi/GBM lightcurves of this event and the best fit AFINO model in the bottom panel. From the lightcurves alone, very clear pulsating structures can be seen in the 25 - 50 keV energy band, and less prominently in the 15 - 25 keV band. This is confirmed by the best-fit AFINO model to the Fourier power spectrum, which identifies a signal with a period of $P$ = 1.8s.

\subsubsection{BN150929278}

This burst trigger was associated with a solar flare from 2015-09-29. The flare was of GOES class M1.4 and began at 06:39 UT, continuing until 06:46 UT. The flare was sourced from AR 12422. 

\begin{figure}
\begin{center}
    \includegraphics[width=8.5cm]{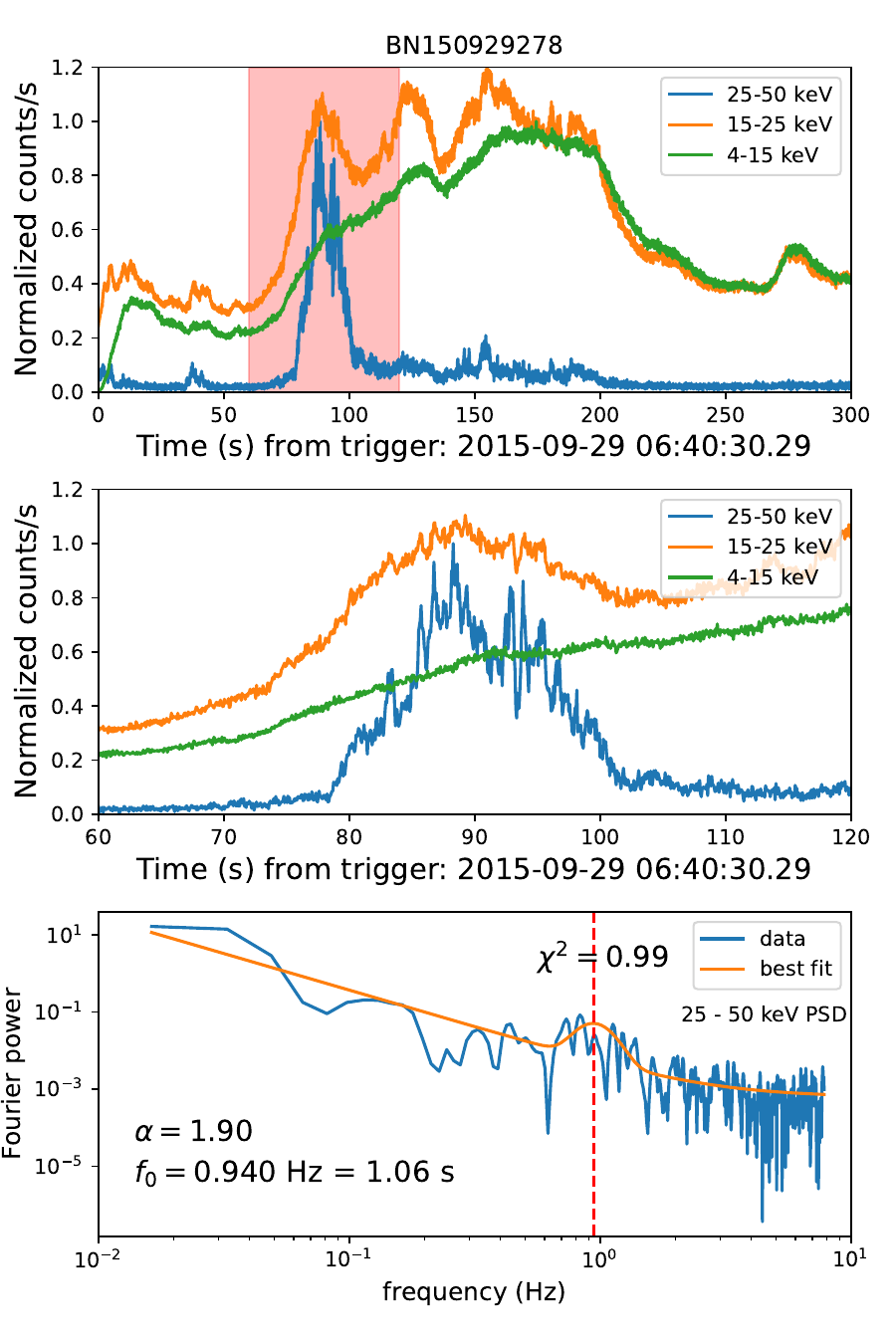}
    \caption{Summary of the BN150929278 solar flare burst. Top: Normalized count rates in the 4-15 keV (green), 15-25 keV (orange), and 25-50 keV (blue) energy ranges at 0.064s cadence. The time series segment where a significant QPP was detected is shaded in red. Center: Zoomed-in lightcurve showing only the time series segment where the QPP was detected, i.e. the red shaded area from the previous panel. Bottom: The Fourier power spectral density (PSD) of the flare time series segment (blue), with the best fit of the QPP-like AFINO model overplotted (orange). Here, a period of $P$ = 1.1s was detected in the 25 -- 50 keV energy band.} 
    \label{summary_bn150929278}
\end{center}
\end{figure}

Figure \ref{summary_bn150929278} shows the Fermi/GBM lightcurves of this flare and the best-fit AFINO model to the Fourier power spectrum. A QPP is detected in the second timeseries segment of this flare, shown by the red highlighted area in the lightcurve. The center panel of Figure \ref{summary_bn150929278} shows a pronounced sequence of pulsations in the 25 - 50 keV band on top of a longer $\sim$ 20s pulse of hard X-rays. The best-fit model in the AFINO analysis indicates a period of $P$ = 1.1s.

\subsubsection{BN220520322}

\begin{figure}
\begin{center}
    \includegraphics[width=8.5cm]{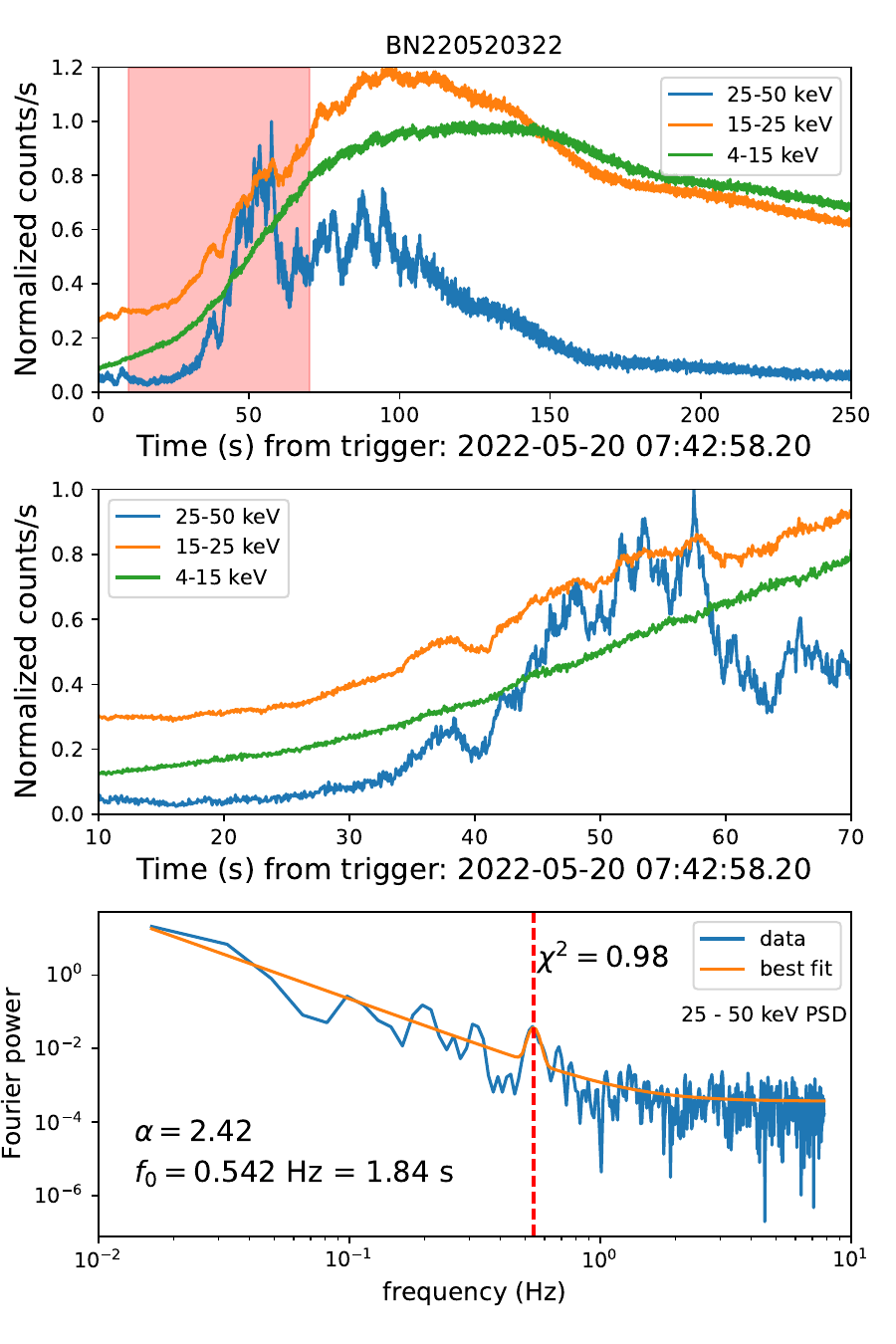}
    \caption{Summary of the BN220520322 solar flare burst. Top: Normalized count rates in the 4-15 keV (green), 15-25 keV (orange), and 25-50 keV (blue) energy ranges at 0.064s cadence. The time series segment where a significant QPP was detected is shaded in red. Center: Zoomed-in lightcurve showing only the time series segment where the QPP was detected, i.e. the red shaded area from the previous panel. Bottom: The Fourier power spectral density (PSD) of the flare time series segment (blue), with the best fit of the QPP-like AFINO model overplotted (orange). Here, a period of $P$ = 1.8s was detected in the 25 -- 50 keV energy band.} 
    \label{summary_bn220520322}
\end{center}
\end{figure}

This burst trigger was associated with a solar flare from 2022-05-20. The flare was of GOES class M3.0 and began at 07:35 UT, continuing until 07:49 UT. AFINO indicates the presence of a periodic signal with $P$ = 1.85s in the 25 -- 50 keV band. This is consistent with the lightcurve shown in the center panel of Figure \ref{summary_bn220520322}, which shows a sequence of periodic bursts between approximately $t$ = 45s and $t$ = 60s.

Figure \ref{summary_bn220520322} also indicates the presence of a prominent, longer period pulsation with a triangular profile occurring throughout the event, most strongly observed in the 25 -- 50 keV energy band. This was not found by AFINO due to the restrictions on the maximum allowed period in the best-fit model, which was 10s. If we remove this period requirement and apply AFINO to the entire 25 -- 50 keV time series between $t$ = 40s and $t$ = 140s, we find strong evidence for a signal with $P$ $\sim$ 9s. Therefore, BN220520322 is a multi-periodic flare, although we note that the two periods do not appear to occur simultaneously; the short-period 1.8s pulsation occurs during the initial hard X-ray peak centered at $t$ = 50s, while the longer 9s pulsation occurs later, from $t$ = 60s onwards. 

\begin{figure}
\begin{center}
\includegraphics[width=8.5cm]{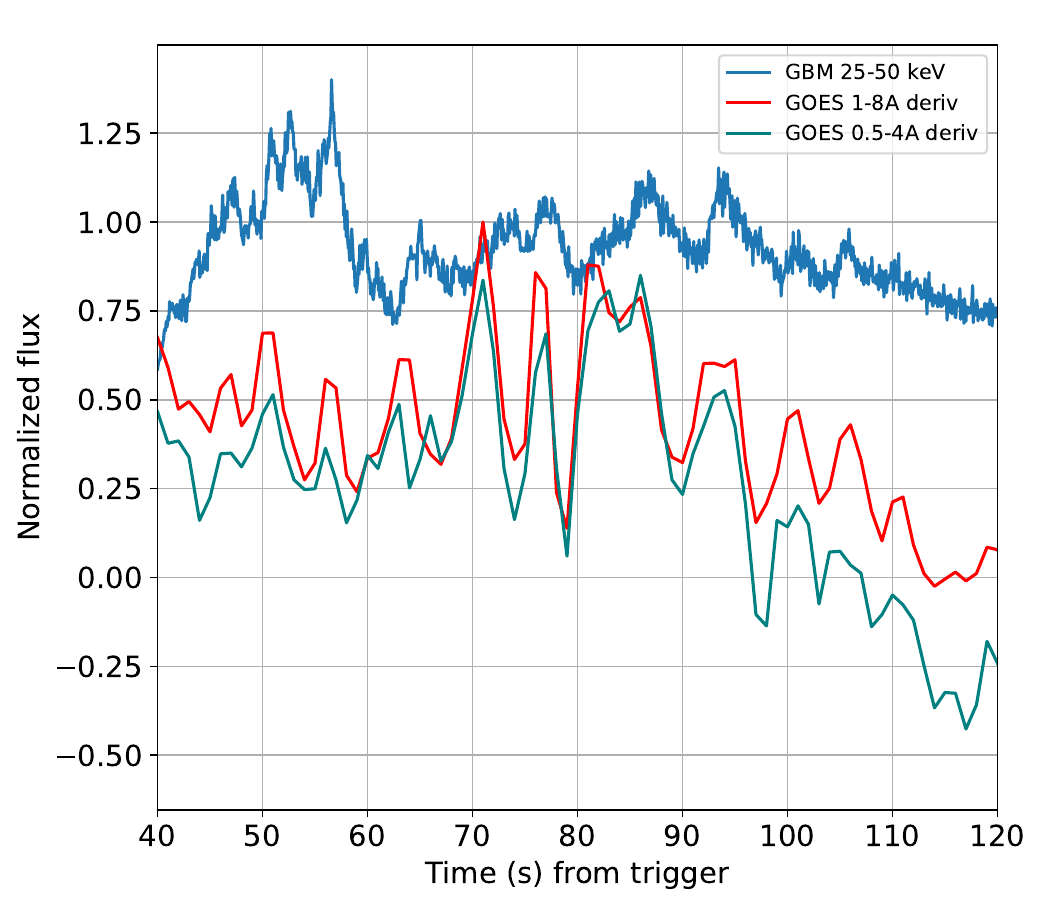}
\caption{Comparison of the GBM 25 -- 50 keV flux for the BN220520322 event with the derivative of the solar soft X-ray flux as measured by the GOES/XRS instrument. The derivative of the GOES long channel (1 -- 8 \AA) is shown in red, while the derivative of the short channel (0.5 -- 4.0 \AA) is shown in magenta. Multiple co-temporal pulses are visible in these lightcurves from approximately $t$ = 70s onwards.}
\label{bn220520_goes_vs_gbm}
\end{center}
\end{figure}

This longer period structure is also observed in the soft X-ray flux observed by GOES at 1s cadence, in both the long (1 -- 8 \AA) and short (0.5 -- 4.0 \AA) wavelength channels. This is illustrated in Figure \ref{bn220520_goes_vs_gbm}, which compares the derivative of the GOES soft X-ray flux with the raw GBM 25 -- 50 keV flux. Hard X-ray solar flare emission is often strongly correlated with the the derivative of thermal soft X-ray flux, a phenomenon known as the Neupert effect \citep{1968ApJ...153L..59N}. In this flare, we see that multiple peaks in the GOES derivative are temporally co-aligned with the peaks in the raw 25 -- 50 keV count rates, particularly from $t$ = 70s after the burst trigger onwards.

\section{Summary}
\label{summary}

We have analysed 1460 solar flares observed by the Fermi/GBM instrument at X-ray energies, in search of short period (<5s) quasi-periodic pulsations. We analyse each flare using the AFINO analysis method, applying it to each flare in a sequence of overlapping temporal windows. To better understand our observational results, we perform tests of our analysis approach on sets of synthetic flare lightcurves, in order to understand our ability to detect such signals and the potential false alarm rates. The main conclusions of this work can be summarized as follows:

\begin{itemize}
\item Based on synthetic flare lightcurves, our analysis method is able to detect short-period, short-lived pulsations in flare lightcurve data (see Figure \ref{simulated_results_fixed_oscillation}). The detection effectiveness depends on both the oscillation frequency and the signal-to-noise ratio, with lower frequencies easier to detect at low SNR. Oscillation frequencies of 0.25 Hz and 0.5 Hz were detectable even when SNR < 1, while oscillation frequencies up to 3 Hz (P = 0.33s) were also detectable by this method provided the SNR was moderate, above $\sim$ 2.

\item We also estimate the false positive rate of this method as a function of detection threshold $\Delta$BIC. This is achieved by applying the method to 1000 flare-like signals that do not contain any oscillations. In addition, we estimate the true positive rate as a function of both $\Delta$BIC and SNR, by applying AFINO to sets of synthetic lightcurves with a fixed SNR. We show that the false alarm rate is more strongly dependent on $\Delta$BIC than the true positive detection rate (see Figure \ref{false_alarm_vs_bic}). This leads us to choose stricter detection thresholds of $\Delta$BIC = 15 and $\Delta$BIC = 20 when searching for pulsations in the real Fermi/GBM data.

\item We do not find strong statistical evidence for a large number of short-period QPP events in the Fermi/GBM burst mode data. Using a strict detection threshold of $\Delta$BIC > 20, we find 13 events in the 4 -- 15 keV energy range, 8 events in the 15 -- 25 keV range, and 5 events in the 25 -- 50 keV range. This compares to our estimated false alarm rate of 0.04\% -- or 6 events -- per energy band. If we relax the detection threshold to $\Delta$BIC > 15, we find 39 events in the 4 -- 15 keV energy range, 17 events in the 15 -- 25 keV range, and 11 events in the 25 -- 50 keV range. Again, this is inline with the false alarm expectation of 30 events, given a false alarm rate of 0.2\%. 

\item Using our analysis of synthetic lightcurves, we can estimate the expected number of detections for a given signal to noise level and base QPP occurrence rate (see Figure \ref{base_rate_plot}). Given the low number of QPP detections in the real GBM X-ray data, we can surmise that either the underlying base occurrence rate of short-period QPPs is very low, or the typical signal to noise ratios of such signals is SNR < 1. This highlights that good signal-to-noise is required in addition to high time resolution when searching for pulsations in solar flares. 

\item Finally, we present a selection of 7 case study QPP events. Each of these flares exhibits a QPP that was detected by AFINO in at least one energy channel with a minimum threshold of $\Delta$BIC > 15. Additionally, each event shows visually compelling evidence of pulsations, indicating a true positive detection and not a false alarm. The periods of these events range from 1 -- 4 s. At least two of these flares -- the M5.1 flare of 2013 October 28 at 04:32 UT, and the M3.0 flare of 2022 May 20 at 07:35 UT -- appear to be multi-periodic events. Observations of multi-periodicity have been relatively rare to date and merit further investigation.

\end{itemize}

This work presents some probable examples of quasi-periodic pulsations in solar flares in the P < 5s range, including two events that may be multi-periodic. However, the overall prevalence of such signatures in flare emission remains unclear. This analysis did not uncover strong evidence for widespread short-period QPPs, due to the difficulty in distinguishing between genuine QPP signals and false alarms when analysing large samples of data. This is partly due to the signal-to-noise ratio available in high cadence solar X-ray data. Future high temporal cadence data with an improved signal to noise ratio are needed in order to resolve this question. Current instruments such as STIX on Solar Orbiter \citep{2020A&A...642A..15K} can provide additional high-cadence observations of QPPs \citep[e.g.][]{2023A&A...671A..79C}, with the advantage of close approaches to the Sun improving the signal to noise ratio. Future focusing optics X-ray telescopes \citep[e.g.][]{2023BAAS...55c.364S}, with intrinsically low background, are a promising avenue for exploring this phenomenon in more detail \citep{2023BAAS...55c.181I}. 

Understanding short-period QPPs is key because they pose questions regarding the fundamental physics of solar flare energy release. Of the many proposed explanations for QPPs, only some can produce periodic behaviour on this timescale in the X-ray regime. Compared to longer period QPPs, short-period events could represent a different population or regime of this phenomenon, strongly implying that multiple different mechanisms produce QPPs depending on the timescale. Confirmed short-period QPPs can therefore provide additional insight and constraints into the timescales of particle acceleration and energy release in flares.  

\bigskip


ARI acknowledges support from the Fermi Guest Investigator program, NASA grant number 80NSSC20K1601. LAH is supported by an ESA Research Fellowship. The authors are grateful to the anonymous referee, who's comments improved the final manuscript.
    

\bibliographystyle{apj}
\bibliography{refs.bib}

\end{document}